\documentclass[useAMS,usegraphicx,usenatbib]{mn2e}                

\usepackage{times}
\usepackage{flushend}

\newcommand{\Msun}{\ensuremath{\,{\rm M}_\odot}}                  
\newcommand{\Rsun}{\ensuremath{\,{\rm R}_\odot}}                  
\newcommand{\Teff}{\ensuremath{T_{\rm eff}}}                      
\newcommand{\logg}{\ensuremath{\log g}}                           
\newcommand{\Mjup}{\ensuremath{\,{\rm M}_{\rm Jup}}}              
\newcommand{\Rjup}{\ensuremath{\,{\rm R}_{\rm Jup}}}              
\newcommand{\Teq}{\ensuremath{T_{\rm eq}^{\,\prime}}}             
\newcommand{\Porb}{\ensuremath{P_{\rm orb}}}                      
\newcommand{\ms}{\,m\,s$^{-1}$}                                   
\newcommand{\mss}{\,m\,s$^{-2}$}                                  
\newcommand{\as}{\ensuremath{^{\prime\prime}}}                    
\newcommand{\am}{\ensuremath{^\prime}}                            
\newcommand{\FeH}{\ensuremath{\left[\frac{\rm Fe}{\rm H}\right]}} 
\newcommand{\rhk}{\ensuremath{\log R^{\prime}_{\rm HK}}}          
\newcommand{\pjup}{\ensuremath{\,\rho_{\rm Jup}}}                 
\newcommand{\psun}{\ensuremath{\,\rho_\odot}}                     
\newcommand{\chir}{\ensuremath{\chi_\nu^{\,2}}}                   
\newcommand{\mc}[1]{\multicolumn{2}{c}{#1}}
\newcommand{\mcc}[1]{\multicolumn{3}{c}{#1}}

\newcommand{\ermcc}[5]{\mcc{\ensuremath{{#1\,^{+#2}_{-#3}}\,^{+#4}_{-#5}}}}

\newcommand{\reff}[1]{#1}

\newcommand{\refff}[1]{#1}

\newcommand{\reffff}[1]{#1}

\title[Transmission spectrum of XO-1\,b]
      {Physical properties and optical-infrared transmission spectrum of the giant planet XO-1\,b}

\author[Southworth et al.]
       {John Southworth\,$^{1}$, J.\ Tregloan-Reed\,$^{2,3}$, A.\ Pinhas\,$^{4}$, N.\ Madhusudhan\,$^{4}$, L.\ Mancini\,$^{5,6,7}$, \newauthor A.\ M.\ S.\ Smith$^{8}$
        \\
        $^{1}$\,Astrophysics Group, Keele University, Staffordshire, ST5 5BG, UK \\
        $^{2}$\,Centro de Astronom\'{\i}a, Universidad de Antofagasta, Avenida U.\ de Antofagasta, 02800, Antofagasta, Chile \\
        $^{3}$\,Carl Sagan Center, SETI Institute, Mountain View, CA 94043, USA \\
        $^{4}$\,Institute of Astronomy, University of Cambridge, Madingley Road, Cambridge CB3 0HA, UK \\
        $^{5}$\,Department of Physics, University of Rome Tor Vergata, Via della Ricerca Scientifica 1, 00133 -- Roma, Italy \\
        $^{6}$\,Max Planck Institute for Astronomy, K\"{o}nigstuhl 17, 69117 -- Heidelberg, Germany \\
        $^{7}$\,INAF -- Astrophysical Observatory of Turin, Via Osservatorio 20, 10025 -- Pino Torinese, Italy \\
        $^{8}$\,Institute of Planetary Research, German Aerospace Center, Rutherfordstrasse 2, 12489 Berlin, Germany
        }

\begin{document} \maketitle 

\begin{abstract}
We present ten high-precision light curves of four transits in the XO-1 planetary system, obtained using $u$, $g$, $r$, redshifted H$\alpha$, $I$ and $z$ filters. We use these to measure the physical properties, orbital ephemeris, and a transmission spectrum of the planet covering the full optical wavelength range. We augment this with published HST/WFC3 observations to construct a transmission spectrum of the planet covering 0.37 to 1.65\,$\mu$m. Our best-fitting model to this spectrum is for a H$_2$/He-rich atmosphere containing water (3.05$\sigma$ confidence), nitrogen-bearing molecules NH$_3$ and HCN (1.5$\sigma$) and patchy cloud (1.3$\sigma$). \reff{We find that adding the optical to the near-infrared data does not lead to more precise constraints on the planetary atmosphere in this case.} We conduct a detailed investigation into the effect of stellar limb darkening on our results, concluding that the choice of limb darkening law and coefficients is unimportant; such conclusions may not hold for other systems so should be reassessed for all high-quality datasets. The planet radius we measure in the $g$-band is anomalously low and should be investigated with future observations at a higher spectral resolution. 
From the measured times of transit we determine an improved orbital ephemeris, calculate a lower limit on the modified stellar tidal quality factor of $Q_\star^{\,\prime} > 10^{5.6}$, and rule out a previously postulated sinusoidal variation in the transit times.
\end{abstract}


\section{Introduction}                                                                                                                                                                \label{sec:intro}

\begin{table*} \centering
\caption{\label{tab:obslog} Log of the observations presented in this work. $N_{\rm obs}$ is
the number of observations, $T_{\rm exp}$ is the exposure time, $T_{\rm dead}$ is the dead time
between exposures, `Moon illum.' is the fractional illumination of the Moon at the midpoint of
the transit, and $N_{\rm poly}$ is the order of the polynomial fitted to the out-of-transit data.
The aperture radii refer to the target aperture, inner sky and outer sky, respectively.}
\setlength{\tabcolsep}{4pt}
\begin{tabular}{lcccccccccccc} \hline
Telescope / & Date of   & Start time & End time &$N_{\rm obs}$&$T_{\rm exp}$&$T_{\rm dead}$& Filter & Airmass &  Moon  & Aperture   &$N_{\rm poly}$& Scatter \\
instrument  & first obs &    (UT)    &   (UT)   &             &     (s)     &     (s)      &        &         & illum. & radii (px) &              & (mmag)  \\
\hline
INT/WFC     & 2010 04 27 & 23:56 & 03:33 &  95 &    100   & 30 & red.\,H$\alpha$ & 1.29 $\to$ 1.00 $\to$ 1.01 & 0.995 & 25 40 60 & \refff{2} & 0.63 \\ 
CAHA/BUSCA  & 2012 05 07 & 22:34 & 03:56 & 107 &    120   & 61 & SDSS $u$        & 1.19 $\to$ 1.01 $\to$ 1.25 & 0.945 & 27 33 60 & 1 & 3.35 \\ 
CAHA/BUSCA  & 2012 05 07 & 22:34 & 03:56 & 107 &    120   & 61 & SDSS $g$        & 1.19 $\to$ 1.01 $\to$ 1.25 & 0.945 & 30 40 70 & 1 & 0.76 \\ 
CAHA/BUSCA  & 2012 05 07 & 22:34 & 03:56 & 107 &    120   & 61 & SDSS $r$        & 1.19 $\to$ 1.01 $\to$ 1.25 & 0.945 & 32 42 80 & 1 & 0.55 \\ 
CAHA/BUSCA  & 2012 05 07 & 23:19 & 03:56 &  92 &    120   & 61 & SDSS $z$        & 1.10 $\to$ 1.01 $\to$ 1.25 & 0.945 & 30 40 70 & 1 & 0.75 \\ 
CAHA/BUSCA  & 2012 05 11 & 21:35 & 03:58 & 142 & 100--120 & 52 & SDSS $u$        & 1.34 $\to$ 1.01 $\to$ 1.31 & 0.595 & 23 33 60 & 1 & 1.44 \\ 
CAHA/BUSCA  & 2012 05 11 & 21:35 & 03:47 & 138 & 100--120 & 52 & SDSS $g$        & 1.34 $\to$ 1.01 $\to$ 1.27 & 0.595 & 30 40 70 & 1 & 0.64 \\ 
CAHA/BUSCA  & 2012 05 11 & 21:35 & 03:58 & 147 & 100--120 & 52 & SDSS $r$        & 1.34 $\to$ 1.01 $\to$ 1.31 & 0.595 & 30 45 80 & 2 & 1.78 \\ 
CAHA/BUSCA  & 2012 05 11 & 21:35 & 03:58 & 145 & 100--120 & 52 & SDSS $z$        & 1.34 $\to$ 1.01 $\to$ 1.31 & 0.595 & 25 40 80 & 2 & 0.72 \\ 
CAHA/1.23m  & 2014 05 26 & 21:39 & 02:58 & 151 & 105--125 & 11 & Cousins $I$     & 1.15 $\to$ 1.01 $\to$ 1.30 & 0.036 & 30 45 80 & 1 & 0.61 \\ 
\hline \end{tabular} \end{table*}

The atmospheric properties of giant planets are an important indicator of the formation and evolution of planets and planetary systems \citep{Madhusudhan++14apj,Mordasini+16apj}. They are also observationally accessible in a large fraction of hot Jupiters (planets with mass $>$0.3\Mjup\ and orbital period $<$10\,d) which transit their host star, via the method of {\it transmission spectroscopy}.

Transmission spectroscopy offers a way of measuring the radius of the planet as a function of wavelength, by determining the transit depth at multiple wavelengths. It is sensitive to the amount of absorption and scattering of starlight passing though the outer atmosphere of the planet whilst it is backlit by its host star. Transmission spectroscopy can be used to detect enhanced opacity due to atomic absorption, molecular absorption and Rayleigh scattering \citep[e.g.][]{Pont+13mn,Madhusudhan++14apj,Fischer+16apj}. This can yield constraints on the chemical composition of the atmosphere, its temperature-pressure structure, and the presence of cloud or haze particles. The first detection of the atmosphere of an extrasolar planet was \reff{due to} sodium absorption in HD\,209458\,b \citep{Charbonneau+02apj}, and extensive results have recently been obtained from both the ground and space \citep[e.g.][]{Nikolov+16apj,Sing+16nat}.

In the near future the NASA {\it James Webb Space Telescope} (JWST) will revolutionise this research area with extensive observations covering wavelengths from 0.6\,$\mu$m to 28\,$\mu$m \citep{Beichman+14pasp,Greene+16apj}. \reff{It is expected to be used to study a significant sample of planets, and by comparison to HST it will achieve much lower Poisson noise, more extensive wavelength intervals, and uninterrupted coverage of individual transits.}

\reff{\citet{Stevenson+16pasp} outlined an {\it Early Release Science} program intended to occur shortly after JWST enters service, in which extensive observations of a small number of transiting planets will be performed using multiple instruments and observing modes. The aims are to allow an assessment of the relative strengths of the observing modes, and to expedite the development of data reduction pipelines for this work. \citeauthor{Stevenson+16pasp} selected 12 transiting planets as promising targets. XO-1 is one of the most suitable targets within this list, with a sky position near the continuous viewing zone of JWST, a host star which is bright ($K_s = 9.53$; \citealt{Skrutskie+06aj}) and inactive ($\rhk = -4.958$; \citealt{Knutson++10apj}), and a planet with an atmospheric scale height (277\,km) suitable for obtaining transmission spectra with a significant signal to noise ratio.} In this work we present a detailed analysis of the XO-1 system, based on new transit light curves in six optical passbands plus published infrared transmission spectroscopy, in order to measure the physical properties of the system, refine the orbital ephemeris, and investigate the atmospheric properties of the planet.

XO-1 was only the eleventh transiting extrasolar planet (TEP) discovered \citep{Mccullough+06apj}, and was found to be a 0.92\Mjup\ and 1.21\Rjup\ planet orbiting a 1.04\Msun\ and 0.94\Rsun\ G1\,V star every 3.94\,d \citep{Me10mn}. Follow-up light curves from \citet{Holman+06apj} were analysed using homogeneous methods by both \citet{Torres++08apj} and \citet{Me08mn,Me09mn,Me11mn}. Occultations (secondary eclipses) were observed by \citet{Machalek+08apj} using the {\it Spitzer Space Telescope}, from which flux ratios of the planet to the star were measured in the four IRAC passbands (3.6, 4.5, 5.8 and 8.0 $\mu$m). \citet{Tinetti+10apj} and \citet{Burke+10apj} presented and studied HST/NICMOS transmission spectroscopy of XO-1\,b, finding evidence for the presence of the molecules H$_2$O, CH$_4$ and CO$_2$. Their results have been questioned by \citet{Gibson++11mn} based on a reanalysis of the same data, and by \citet{Deming+13apj} based on new HST/WFC3 transmission spectroscopy. \citeauthor{Deming+13apj} found evidence for water in the transmission spectrum of XO-1\,b, a conclusion also reached by \citet{Tsiaras+18aj}.

In addition to the works cited above, transit light curves have been presented by \citet{Vanko+09conf,Caceres+09aa,Raetz+09an} and \citet{Sada+12pasp}, spectroscopic analyses of the host star have been performed by \citet{Ammler+09aa,Torres+12apj,Mortier+13aa} and \citet{Teske+14apj}, and the orbital eccentricity has been constrained to be less than 0.29 by \citet{MadhusudhanWinn09apj} and \citet{Pont+11mn}. Most recently, \citet{Bonomo+17aa} presented new radial velocity measurements from which they constrained the eccentricity to be less than 0.019 to $1\sigma$ and 0.043 to $2\sigma$.

High-resolution imaging of TEP host stars is an important part of determining the physical properties of the system \citep[e.g.][]{Evans++16apj}. Lucky Imaging of the XO-1 system was presented by \citet{Wollert+15aa}, who found no nearby stars \reffff{less than 3.97, 4.85, 5.79 and 6.46 mag fainter than XO-1\,A} (5$\sigma$ detection limits) in the $z^\prime$ band at distances of 0.25\as, 0.5\as, 1.0\as\ and 2.0\as, respectively.

\reff{The outline of this paper is as follows. Section\,\ref{sec:obs} presents our new observations of XO-1, which are analysed in Section\,\ref{sec:lc} alongside published light curves. The results are used to measure the physical properties of the system in Section\,\ref{sec:absdim}. Section\,\ref{sec:porb} presents an improved orbital ephemeris and a search for transit timing variations. The transmission spectrum of XO-1\,b is obtained and analysed in Section\,\ref{sec:transspec}, after which the paper is concluded in Section\,\ref{sec:summary}.}


\section{Observations and data reduction}                                                                                                                                               \label{sec:obs}

\begin{figure*} \includegraphics[width=\textwidth,angle=0]{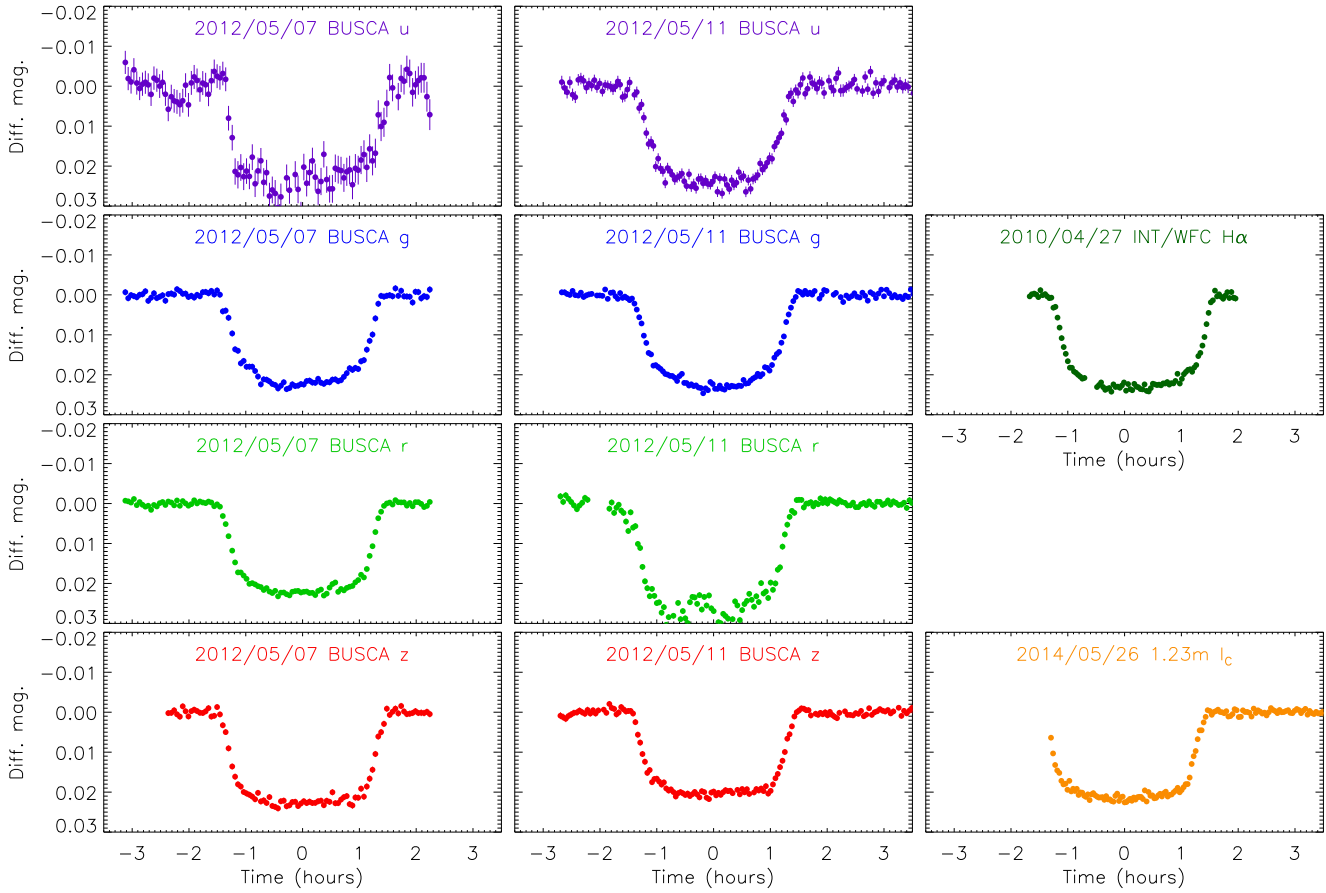}
\caption{\label{fig:lcall} The new light curves presented in this work.
Times are given relative to the midpoint of each transit. The date,
instrument and filter used is indicated.} \end{figure*}

Two transits of XO-1 were observed using the BUSCA four-band imaging photometer on the 2.2\,m telescope at Calar Alto, Spain. We selected Sloan $u$, $g$, $r$ and $z$ filters \citep{Fukugita+96aj} from the Calar Alto filter database, which with BUSCA yield a circular field of view approximately 5.8\am\ in diameter. We were not able to obtain good photometry simultaneously in the bands with the lowest and highest counts ($u$ and $r$ respectively), \reff{because the four arms of BUSCA cannot be operated at different focus levels or exposure times. We therefore optimised for the $r$-band on the first night and the $u$-band on the second night.} The $u$-band light curve from 2012/05/07 therefore has a large scatter due to low flux levels, and the $r$- band light curve from 2012/05/11 displays systematic effects due to being near the saturation level of the CCD. An observing log is given in Table\,\ref{tab:obslog} and further details of our approach with BUSCA can be found in \citet{Me+12mn2}.

One transit of XO-1 was observed with the Isaac Newton Telescope (INT) and Wide Field Camera (WFC) on La Palma, Spain. We used CCD 4, as this is the one on the optical axis, and a redshifted H$\alpha$ filter (ING filter\footnote{{\tt http://catserver.ing.iac.es/filter/list.php?\\instrument=WFC}} \#226, central wavelength 689\,nm, FWHM 10\,nm) rather than a wide-band filter in order to limit the amount of defocussing used. We were not able to autoguide as the guide CCD is in the same focal plane as the science instrument.

One transit of XO-1 was obtained with the 1.23\,m telescope at Calar Alto, using a Cousins $I$ filter. The transit ingress was missed due to technical problems, but the light curve is otherwise excellent.

The data were reduced using the {\sc defot} pipeline \citep{Me+09mn,Me+14mn}, which depends on the NASA {\sc astrolib} library\footnote{{\tt http://idlastro.gsfc.nasa.gov/}} {\sc idl}\footnote{{\tt http://www.harrisgeospatial.com/SoftwareTechnology/ IDL.aspx}} implementation of the {\sc aper} routine from {\sc daophot} \citep{Stetson87pasp}. Software apertures were placed by hand and their radii chosen to minimise the scatter in the final light curve. \reff{The apertures were shifted to account for telescope pointing wander, which was measured by cross-correlating each image with a reference image.} We did not perform bias or flat-field calibrations as these had little effect on the final light curves beyond a slight increase in the scatter of the data.

A differential-magnitude light curve of XO-1 was generated for each observing sequence versus an ensemble comparison star containing the weighted flux sum of the good comparison stars. A polynomial was also fitted to the observations outside transit and subtracted to rectify the final light curve to zero differential magnitude. \reff{The order of the polynomial was chosen to be the lowest which gave a good fit to the out-of-transit data.} The coefficients of the polynomial and the weights of the comparison stars were simultaneously optimised to minimise the scatter in the datapoints outside eclipse. The observational errors were then scaled so each transit had a reduced $\chi^2$ of $\chir = 1.0$ versus a best-fitting model calculated with the {\sc jktebop} code (see below). Table\,\ref{tab:obslog} includes the polynomial order and the $rms$ of the residuals versus the best fit for each light curve. The final data are shown in Fig.\,\ref{fig:lcall} and listed in Table\,\ref{tab:lcdata}. The times given refer to the midpoint of the exposure on the BJD/TDB timescale \citep{Eastman++10pasp}.


\section{Light curve analysis}                                                                                                                                                           \label{sec:lc}

\subsection{Approach}

We modelled the available transit light curves of XO-1 using the {\sc jktebop}\footnote{{\sc jktebop} is written in {\sc fortran77} and the source code is available at {\tt http://www.astro.keele.ac.uk/jkt/codes/jktebop.html}} code \citep[][and references therein]{Me13aa} and the formalism of the {\it Homogeneous Studies} project \citep[see][and references therein]{Me12mn}. The fitted parameters were as follows.
\begin{itemize}
\item The sum and ratio of the fractional radii of the two components, $r_{\rm A} + r_{\rm b}$ and $k = \frac{r_{\rm b}}{r_{\rm A}}$, where the fractional radii are the absolute radii in units of the semimajor axis: $r_{\rm A,b}= \frac{R_{\rm A,b}}{a}$. These combinations of parameters were chosen because they are only weakly correlated.
\item The orbital inclination, $i$.
\item A time of mid-transit, $T_0$.
\item The coefficients of a polynomial of differential magnitude versus time. The polynomial order for each light curve is given in Table\,\ref{tab:obslog}. Whilst the fitted polynomials were removed at the data-reduction stage, their inclusion in the {\sc jktebop} fit is necessary to propagate their uncertainties into the measured photometric parameters.
\item One or two limb darkening (LD) coefficients, depending on the solution performed.
\end{itemize}

The orbital period was held fixed in each solution, because the uncertainty in its value was utterly negligible for this analysis. We also enforced orbital circularity in the absence of evidence for an eccentric orbit \citep[see discussion in][]{Anderson+12mn}.

\begin{table} \centering \caption{\label{tab:lcdata} The
first line of each of the light curves presented in this
work. The full dataset will be made available at the CDS.}
\begin{tabular}{llcrr} \hline
Telescope / & Filter & BJD(TDB)   & Diff.\ mag. & Uncertainty \\
instrument  &        & $-$2400000 &             &             \\
\hline
INT/WFC   & rH$\alpha$ & 55314.497407 & $ $0.0003821 & 0.0006442 \\
BUSCA     & $u$        & 56055.445999 & $-$0.0059505 & 0.0028984 \\
BUSCA     & $g$        & 56055.445999 & $-$0.0006606 & 0.0007777 \\
BUSCA     & $r$        & 56055.445999 & $-$0.0007141 & 0.0005374 \\
BUSCA     & $z$        & 56055.477499 & $ $0.0002283 & 0.0007324 \\
BUSCA     & $u$        & 56059.405206 & $-$0.0010300 & 0.0015896 \\
BUSCA     & $g$        & 56059.405206 & $-$0.0006314 & 0.0006988 \\
BUSCA     & $r$        & 56059.405206 & $-$0.0018004 & 0.0007984 \\
BUSCA     & $z$        & 56059.405206 & $ $0.0008813 & 0.0008654 \\
CAHA123   & $I$        & 56804.408021 & $ $0.0063808 & 0.0005926 \\
\hline \end{tabular} \end{table}

We performed {\sc jktebop} solutions using each of four two-parameter LD `laws': quadratic, square-root, logarithmic and cubic \citep[see][]{Me08mn}. We furthermore calculated solutions with both LD coefficients fixed at theoretical values, the linear coefficient fitted and the nonlinear coefficient fixed, and both coefficients fitted. The theoretical LD coefficients were obtained by bilinearly interpolating\footnote{Bilinear interpolation was performed using the {\sc jktld} code at: {\tt http://www.astro.keele.ac.uk/jkt/codes/jktld.html}} in tabulated predictions to the host star's \reff{measured} effective temperature (\Teff) and surface gravity (\logg). We considered multiple sources of theoretical coefficients \citep{Vanhamme93aj,Claret00aa,ClaretHauschildt03aa,Claret04aa2} and averaged their predictions when necessary.

Least-squares best fits were obtained using the Levenberg-Marquardt method \citep{Marquardt63} as implemented in the {\sc mrqmin} routine \citep{Press+92book}. The uncertainties in the fitted parameters were estimated using both Monte Carlo and residual-permutation solutions \citep[see][for further details]{Me08mn}, and the larger errorbar was retained for each measured parameter. Uncertainties were further inflated to account for any scatter in the measured values of a parameter from the solutions using different approaches to the inclusion of LD. Tables of results for each light curve can be found in the Supplementary Information. The measured photometric parameters are given in Table\,\ref{tab:lcfit}.

\subsection{Our new data}

\begin{table*} \centering \caption{\label{tab:lcfit} Parameters of
the fit to the light curves of XO-1 from the {\sc jktebop} analysis.
The final weighted-mean parameters are given in bold.}
\begin{tabular}{l r@{\,$\pm$\,}l r@{\,$\pm$\,}l r@{\,$\pm$\,}l r@{\,$\pm$\,}l r@{\,$\pm$\,}l}
\hline
Source & \mc{$r_{\rm A}+r_{\rm b}$} & \mc{$k$} & \mc{$i$ ($^\circ$)} & \mc{$r_{\rm A}$} & \mc{$r_{\rm b}$} \\
\hline
BUSCA $u$                     & 0.1048 & 0.0063 & 0.1372 & 0.0032 & 87.22 & 1.17 & 0.0923 & 0.0053 & 0.01253 & 0.00101 \\
BUSCA $g$                     & 0.0993 & 0.0024 & 0.1289 & 0.0014 & 89.07 & 0.74 & 0.0880 & 0.0020 & 0.01133 & 0.00038 \\
BUSCA $r$                     & 0.1002 & 0.0024 & 0.1331 & 0.0012 & 88.82 & 0.81 & 0.0884 & 0.0021 & 0.01177 & 0.00037 \\
BUSCA $z$                     & 0.1014 & 0.0027 & 0.1327 & 0.0015 & 88.50 & 0.81 & 0.0895 & 0.0023 & 0.01188 & 0.00038 \\
INT/WFC rH$\alpha$            & 0.0977 & 0.0015 & 0.1339 & \refff{0.0027} & 89.88 & 0.62 & 0.0862 & 0.0013 & 0.01154 & 0.00022 \\
1.23\,m $I_C$                 & 0.1012 & 0.0029 & 0.1334 & 0.0013 & 88.52 & 0.58 & 0.0893 & 0.0025 & 0.01191 & 0.00042 \\[3pt]
\citet{Holman+06apj} FLWO     & 0.1009 & 0.0023 & 0.1321 & 0.0012 & 88.56 & 0.66 & 0.0891 & 0.0019 & 0.01177 & 0.00033 \\
\citet{Holman+06apj} Palomar  & 0.0953 & 0.0043 & 0.1265 & 0.0028 & 89.99 & 1.03 & 0.0846 & 0.0037 & 0.01070 & 0.00063 \\
\citet{Caceres+09aa} SofI     & 0.1018 & 0.0025 & 0.1324 & 0.0021 & 88.47 & 0.47 & 0.0899 & 0.0022 & 0.01191 & 0.00036 \\
\citet{Caceres+09aa} ISAAC    & 0.0978 & 0.0038 & 0.1321 & 0.0018 & 89.81 & 0.95 & 0.0863 & 0.0033 & 0.01140 & 0.00063 \\
\citet{Sada+12pasp}           & 0.1082 & 0.0110 & 0.1297 & 0.0063 & 87.92 & 1.86 & 0.0958 & 0.0092 & 0.01242 & 0.00176 \\[3pt]
Final results&{\bf0.0997}&{\bf0.0008}&{\bf0.1325}&{\bf0.0008}&{\bf88.84}&{\bf0.22}&{\bf0.0880}&{\bf0.0007}&{\bf0.01166}&{\bf0.00012}\\
\hline \end{tabular} \end{table*}

\begin{figure} \includegraphics[width=\columnwidth,angle=0]{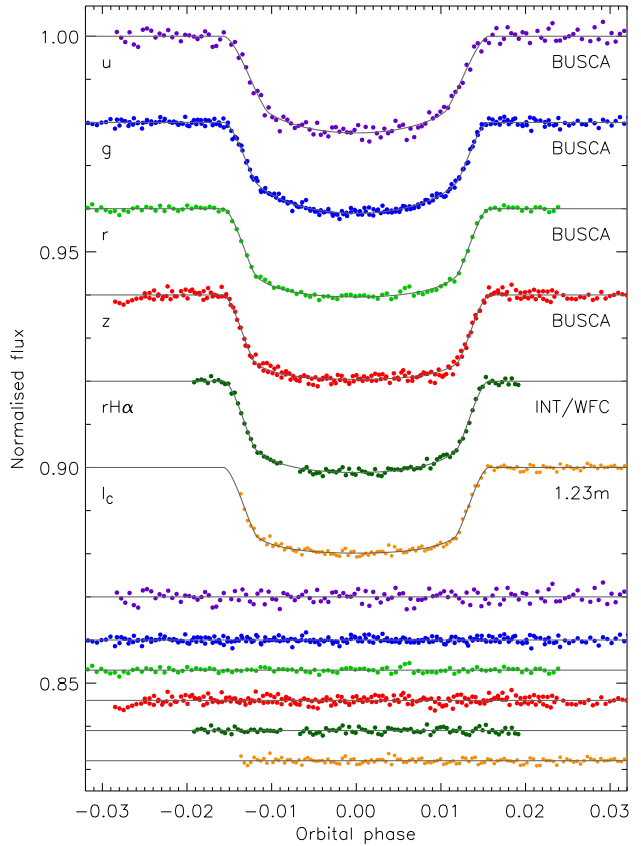}
\caption{\label{fig:lcfit} {\sc jktebop} best fits to our phased light curves of XO-1.
The data are shown as filled circles colour-coded consistently with Fig.\,\ref{fig:lcall}.
The best fits are shown as grey lines. The residuals are offset to appear at the base of
the figure. Labels give the passband and source for each dataset. The polynomial baseline
functions have been subtracted from the data before plotting.} \end{figure}

The data from the two transits of XO-1 observed with BUSCA were collected into one light curve for each filter, and each was modelled with {\sc jktebop} (see Fig.\,\ref{fig:lcfit}). We made two exceptions to this approach: the $u$-band data from 2012/05/07 were ignored because the low flux levels caused a large scatter, and the $r$-band data from 2012/05/11 were rejected because they suffer from saturation effects. We found that the $g$-band light curves are in excellent agreement with each other ($\chir = 1.02$ when the individual light curves have $\chir = 1.0$). However, the $z$-band light curves are not ($\chir = 1.56$), as can be seen in Fig.\,\ref{fig:lcall}. Our resulting parameters for the $z$-band are therefore roughly the average for the two datasets, and are in fact in good agreement with the results from other light curves.

In all cases except BUSCA $u$, we adopt the results from {\sc jktebop} models with the linear LD coefficient fitted and the nonlinear LD coefficient fixed, as these agree very well both between different LD laws and between different datasets. For BUSCA $u$ we found that the data were unable to support fitting for even one LD coefficient, so we adopt the results obtained with both coefficients fixed. LD coefficients are not available for the redshifted H$\alpha$ filter. We therefore used those for the Johnson $R$ filter, which has a similar central wavelength (0.67\,$\mu$m; \citealt{Johnson66}). The effect of the difference in passband on the LD coefficients is expected to be smaller than the intrinsic uncertainty of the coefficients, as judged from the variation in predictions for the same filters from different sources.

\subsection{Reanalysis of published data}

\begin{figure} \includegraphics[width=\columnwidth,angle=0]{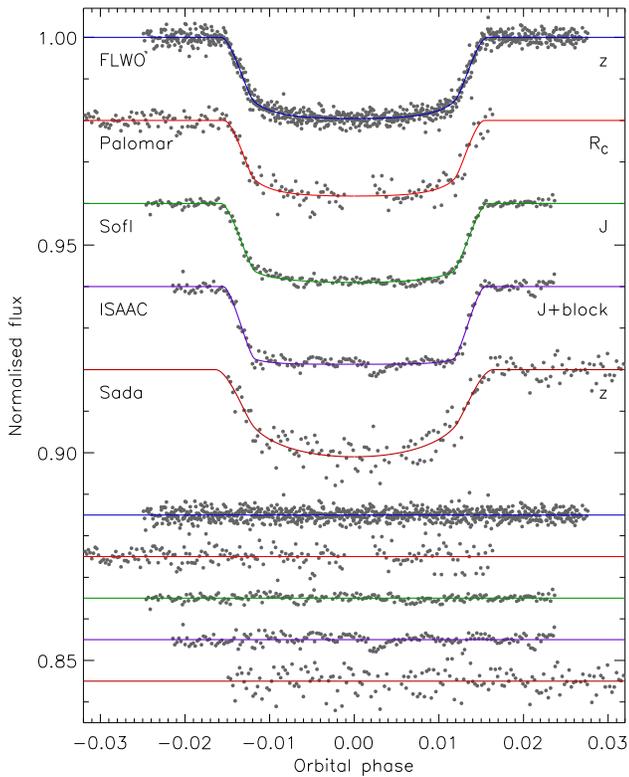}
\caption{\label{fig:lcfit:pub} {\sc jktebop} best fits to published light curves of XO-1.
The data are shown as filled circles and the best fits as grey lines. The residuals are
offset to appear at the base of the figure. Labels give the passband and source for each
dataset. The polynomial baseline functions have been subtracted from the data before
plotting.} \end{figure}

The literature for XO-1 includes several light curves of a quality sufficient for inclusion in the current work. We have obtained these data and modelled them using the same methods as for our own observations. The results are included in Table\,\ref{tab:lcfit} and are discussed below.

\citet{Holman+06apj} presented light curves of two transits of XO-1 obtained with the FLWO 1.2\,m telescope and KeplerCam in the $z$-band, and one transit observed using the Palomar 1.5\,m in the $R$ band. According to the webpage for this facility\footnote{{\tt http://www.astro.caltech.edu/palomar/observer/\\P60observers.html}} this corresponds to a Kron-Cousins $R$ band. Both datasets have been analysed in the past by the first author \citep{Me08mn} but were reanalysed with the modification that a first-order polynomial was applied to each transit, an option added to {\sc jktebop} since the previous analysis \citep[see][]{Me+14mn}. The best fits are shown in  Fig.\,\ref{fig:lcfit:pub} and were each obtained with one fitted and one fixed LD coefficient.

\citet{Caceres+09aa} published observations of four transits of XO-1, all obtained at near-infrared wavelengths. We ignored their Run A due to the large systematic errors visible in the data, and their Run C due to the patches of very high scatter during transit. We therefore analysed their Run B, which was obtained using NTT/SofI in the $J$-band, and their Run D, observed using VLT/ISAAC in the $J$-band but with a blocking filter to remove flux from a red leak in the $J$ filter. Both runs were obtained at high cadence, with integration times of 0.8\,s and 0.08\,s respectively and very low dead times. We therefore binned the light curves by factors of 100 and 1000, respectively, to yield a sampling rate of approximately 80\,s in both cases. Whereas the SofI data could be satisfactorily modelled with one fitted LD coefficient, we had to fix both to obtain an acceptable solution of the ISAAC observations. In both cases we included a second-order polynomial to model the baseline brightness of the system.

\citet{Sada+12pasp} observed one transit in the $z$-band using a 0.5\,m telescope at Kitt Peak National Observatory. The ingress was missed and the data have an expectedly large scatter of 2.9\,mmag, but we ran the usual {\sc jktebop} solutions in order to determine whether this dataset can provide results worth including on our analysis. We allowed for a second-order polynomial baseline function.

\subsection{Combined results}

\begin{figure} \includegraphics[width=\columnwidth,angle=0]{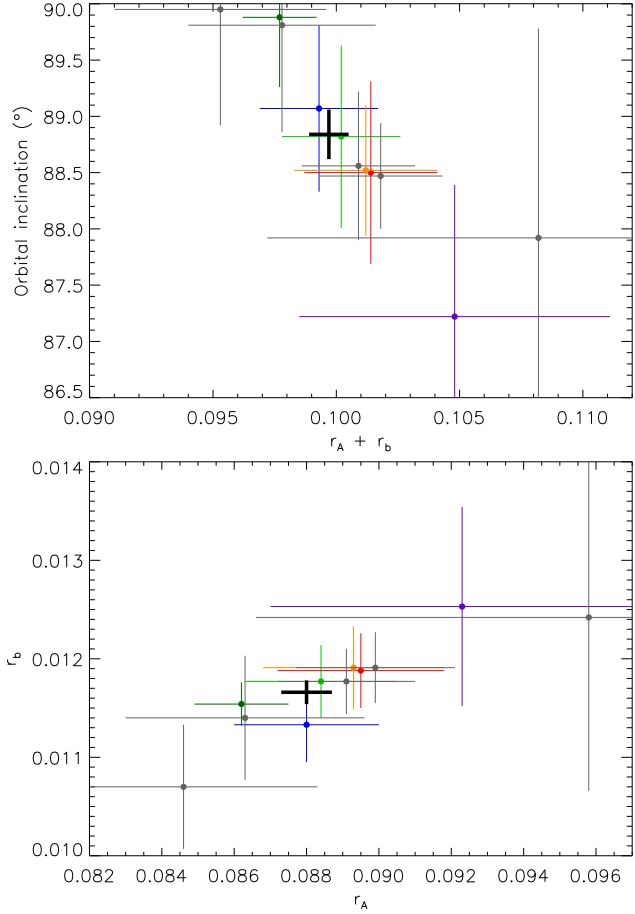}
\caption{\label{fig:lcfit:compare} Plots of the measured values of the sum of the
fractional radii versus orbital inclination (top) and of the two fractional radii
(bottom). In each case the coloured points represent the light curves presented
in this work, with colour-coding the same as in Fig.\,\ref{fig:lcall}, grey points
show results for literature light curves, and the bold black lines indicate the
weighted mean value.} \end{figure}

The measured photometric parameters are given in Table\,\ref{tab:lcfit} and show a good agreement between light curves. We calculated the weighted mean value for each measured parameter for use in the next section. The \chir\ value of the individual values versus the weighted mean is good for $r_{\rm A}$, $r_{\rm b}$ and $i$ (0.6, 0.6 and 0.9, respectively) but less so for $k$ (1.8). This could be caused by residual systematic errors and/or by a true astrophysical signal (i.e.\ a non-flat transmission spectrum; see \citealt{MeEvans16mn}). We have multiplied the uncertainty in the weighted mean of $k$ by $\sqrt{1.8}$ to account for this.

Casual inspection of Table\,\ref{tab:lcfit} suggests that correlations exist between several of the photometric parameters. Such correlations are widely known \citep[e.g.][]{Me08mn,Pal08mn,Carter+08apj} and must be accounted for in the uncertainties of the parameter measurements. In Fig.\,\ref{fig:lcfit:compare} we illustrate two of these correlations: between $r_{\rm A}+r_{\rm b}$ and $i$, and between $r_{\rm A}$ and $r_{\rm b}$. The former arises because $r_{\rm A}+r_{\rm b}$ and $i$ together determine the duration of the transit \citep{SeagerMallen03apj}, and the latter occurs as $k = \frac{r_{\rm b}}{r_{\rm A}}$ is much better-determined than \reff{either} $r_{\rm A}$ or $r_{\rm b}$. It is clear from Fig.\,\ref{fig:lcfit:compare} that the correlations are greatly attenuated using the high-quality light curves presented here, and that the errorbars in Table\,\ref{tab:lcfit} are not underestimated. For reference, the linear Pearson correlation coefficients are $-0.89$ and $+0.90$, respectively.


\section{Physical properties}                                                                                                                                                        \label{sec:absdim}

\begin{table} \centering \caption{\label{tab:spec} Spectroscopic properties
of XO-1\,A given in the literature. Asterisks denote errorbars which include
statistical but not systematic uncertainties.
\newline {\bf References:}
(1) \citet{Mccullough+06apj}; (2) \citet{Ammler+09aa}; (3) \citet{Torres+12apj};
(4) \citet{Mortier+13aa}; (5) \citet{Teske+14apj}; (6) \citet{Brewer+16apjs}.
}
\begin{tabular}{r@{\,$\pm$\,}l r@{\,$\pm$\,}l r@{\,$\pm$\,}l l}
\hline
\mc{\Teff\ (K)} & \mc{\FeH\ (dex)} & \mc{\logg\ (c.g.s.)} & Ref \\
\hline
5750 & 13$^*$ &    0.015 & 0.040$^*$ & 4.53 & 0.065$^*$ & 1 \\
5754 & 42     & $-$0.01  & 0.05      & 4.61 & 0.05      & 2 \\
5738 & 65     & $-$0.06  & 0.05      & 4.50 & 0.01      & 3 \\
5754 & 42     & $-$0.01  & 0.05      & 4.61 & 0.05      & 4 \\
5695 & 26     & $-$0.11  & 0.06      & 4.42 & 0.12      & 5 \\
5729 & 25$^*$ & $-$0.07  & 0.010$^*$ & 4.49 & 0.028$^*$ & 6 \\
\hline
\multicolumn{6}{l}{Adopted parameters:} \\
5740 & 50 & $-$0.03 & 0.05  &    \mc{ }    &   \\
\hline \end{tabular} \end{table}


\begin{table} \centering \caption{\label{tab:absdim} Derived physical properties of
the XO-1 system \refff{from this work compared to those from \citet{Burke+10apj}}.
When measurements are accompanied by two errorbars, the first refers to the
statistical uncertainties and the second to the systematic uncertainties.}
\begin{tabular}{l r@{\,$\pm$\,}c@{\,$\pm$\,}l r@{\,$\pm$\,}l}
\hline
Parameter & \mcc{This work} & \mc{\refff{\citet{Burke+10apj}}} \\
\hline
$M_{\rm A}$    (\Msun) & 1.018    & 0.028    & 0.034     & ~~~~~~\refff{1.027} & \refff{0.06}  \\
$R_{\rm A}$    (\Rsun) & 0.930    & 0.011    & 0.010     & \refff{ 0.94} & \refff{0.02}  \\
$\log g_{\rm A}$ (cgs) & 4.509    & 0.009    & 0.005     & \refff{ 4.50} & \refff{0.01}  \\
$\rho_{\rm A}$ (\psun) & \mcc{$1.265 \pm 0.030$}         & \refff{ 1.23} & \refff{0.03}  \\[2pt]
$M_{\rm b}$    (\Mjup) & 0.907    & 0.022    & 0.020     & \refff{ 0.92} & \refff{0.08}  \\
$R_{\rm b}$    (\Rjup) & 1.199    & 0.017    & 0.013     & \refff{ 1.21} & \refff{0.03}  \\
$g_{\rm b}$     (\mss) & \mcc{$15.65 \pm  0.40$}         & \refff{ 15.5} & \refff{1.1 }  \\
$\rho_{\rm b}$ (\pjup) & 0.492    & 0.018    & 0.005     & \refff{ 0.48} & \refff{0.04}  \\[2pt]
\Teq\              (K) & \mcc{$1204 \pm   11$}           &    \mc{}      \\
$a$               (AU) & 0.04914  & 0.00045  & 0.00054   & \refff{0.049} & \refff{0.001} \\
Age              (Gyr) & \ermcc{1.1}{1.2}{1.1}{0.9}{1.0} &    \mc{}      \\
\hline \end{tabular} \end{table}

We used the results of the photometric analysis from the previous section to obtain the full physical properties of the system. This process also required knowledge of the spectroscopic properties of the host star (effective temperature \Teff\ and metallicity \FeH) which are summarised in Table\,\ref{tab:spec}, and the stellar orbital velocity amplitude, $K_{\rm A} = 115.3 \pm 1.8$\ms\ \citep{Bonomo+17aa}. As the necessary additional constraint, we used tabulated predictions from each one of five sets of theoretical stellar models \citep{Claret04aa,Demarque+04apjs,Pietrinferni+04apj,Vandenberg++06apjs,Dotter+08apjs}.

We then estimated the value of the velocity amplitude of the planet, $K_{\rm b}$ and calculated the physical properties of the system using this and the measured quantities. We iteratively adjusted $K_{\rm b}$ to optimise the agreement between the calculated $\frac{R_{\rm A}}{a}$ and the measured $r_{\rm A}$, and between the \Teff\ and that predicted by the stellar models for the observed \FeH\ and calculated stellar mass ($M_{\rm A}$). We did this for ages from 0.1\,Gyr to 20\,Gyr in steps of 0.1\,Gyr, from which we identified the overall best fit and age of the system \citep[see][]{Me09mn}. This process was undertaken for each of the five sets of tabulated theoretical model predictions, and the final parameters were taken to be the median of the five different possibilities arising from this repeated analysis.

We propagated the statistical errors in all input parameters using a perturbation analysis, and added all contributions in quadrature for each output parameter. We estimated the systematic uncertainties, which are incurred by the use of theoretical stellar models, by taking the maximum deviation between the final parameter value and the individual values obtained using the different sets of tabulated predictions.

The measured physical properties of the XO-1 system are given in Table\,\ref{tab:absdim}. The mass, radius, gravity and density of the star are denoted by: $M_{\rm A}$, $R_{\rm A}$, $\log g_{\rm A}$ and $\rho_{\rm A}$; and of the planet by $M_{\rm b}$, $R_{\rm b}$, $g_{\rm b}$ and $\rho_{\rm b}$, respectively. \refff{Our results are in good agreement with all previously published measurements}. Our measured $r_{\rm A}$ is equivalent to a relatively large stellar density, which means that the best-fitting theoretical star is near the zero-age main sequence. We therefore see a significant systematic uncertainty in our results caused by edge effects in the model tabulations, and by the intrinsic variation in how different stellar evolution codes initialise their stellar models. \refff{Table\,\ref{tab:absdim} also includes a comparison between our measurements and those of \citet{Burke+10apj}, which are in very good agreement.}

This young age is surprising because it is not supported by other age indicators such as chromospheric activity and rotational velocity. \citet{Knutson++10apj} observed the cores of the Ca\,II H \& K lines, finding a small core emission due to stellar activity. They measured an activity index of $\log R^\prime_{\rm HK} = -4.958$, which indicates that it is a relatively inactive star. The calibration of \citet{MamajekHillenbrand08apj} points to an age of roughly 6\,Gyr with an uncertainty of perhaps 0.05\,dex due to astrophysical scatter, and unknown uncertainties due to the $\log R^\prime_{\rm HK}$ value (which is not supplied with an errorbar) and activity cycles on XO-1\,A (because we only have one measurement of $\log R^\prime_{\rm HK}$). One possible solution to this conflict is inaccuracies in theoretical models \citep[e.g.][]{Maxted++15aa2}, with perhaps a small contribution from an orbital eccentricity which is large enough to affect the measured $r_{\rm A}$ but small enough to hide in the available radial velocity measurements.


\section{Transit timing analysis}                                                                                                                                                      \label{sec:porb}

\begin{table*} \begin{center}
\caption{\label{tab:tmin} Times of minimum light and their residuals versus the ephemerides derived in this work.}
\begin{tabular}{l l r r l} \hline
Time of minimum & Uncertainty & Cycle & Residual & Reference \\
(BJD/TDB) & (d) & number & (d) &       \\
\hline
2453127.03924 & 0.00580 &  $-$555.0 & $ $0.00167 & \citet{Wilson+06pasp} \\                                 
2453150.68624 & 0.01060 &  $-$549.0 & $-$0.00036 & \citet{Wilson+06pasp} \\                                 
2453154.62574 & 0.00260 &  $-$548.0 & $-$0.00236 & \citet{Wilson+06pasp} \\                                 
2453158.56704 & 0.00340 &  $-$547.0 & $-$0.00257 & \citet{Wilson+06pasp} \\                                 
2453162.51444 & 0.00250 &  $-$546.0 & $ $0.00333 & \citet{Wilson+06pasp} \\                                 
2453166.45124 & 0.00250 &  $-$545.0 & $-$0.00138 & \citet{Wilson+06pasp} \\                                 
2453170.39244 & 0.00370 &  $-$544.0 & $-$0.00168 & \citet{Wilson+06pasp} \\                                 
2453229.51504 & 0.00450 &  $-$529.0 & $-$0.00165 & \citet{Wilson+06pasp} \\                                 
2453237.40504 & 0.00320 &  $-$527.0 & $ $0.00534 & \citet{Wilson+06pasp} \\                                 
2453241.34174 & 0.00670 &  $-$526.0 & $ $0.00054 & \citet{Wilson+06pasp} \\                                 
2453808.91774 & 0.00110 &  $-$382.0 & $-$0.00012 & \citet{Mccullough+06apj} \\                              
2453875.92321 & 0.00047 &  $-$365.0 & $-$0.00023 & This work (Palomar data from \citealt{Holman+06apj}) \\  
2453879.86474 & 0.00110 &  $-$364.0 & $-$0.00021 & \citet{Holman+06apj} \\                                  
2453883.80638 & 0.00018 &  $-$363.0 & $-$0.00007 & This work (FLWO data from \citealt{Holman+06apj}) \\     
2453887.74746 & 0.00015 &  $-$362.0 & $-$0.00049 & This work (FLWO data from \citealt{Holman+06apj}) \\     
2453887.74774 & 0.00060 &  $-$362.0 & $-$0.00021 & B.\ Gary (AXA) \\                                        
2453911.39781 & 0.00049 &  $-$356.0 & $ $0.00083 & J.\ Ohlert (TRESCA) \\                                   
2454171.53332 & 0.00170 &  $-$290.0 & $-$0.00298 & \citet{Raetz+09an} \\                                    
2454214.89274 & 0.00090 &  $-$279.0 & $-$0.00011 & B.\ Gary (AXA) \\                                        
2454218.83405 & 0.00114 &  $-$278.0 & $-$0.00030 & \citet{Caceres+09aa} \\                                  
2454222.77623 & 0.00023 &  $-$277.0 & $ $0.00037 & This work (SofI data from \citealt{Caceres+09aa}) \\     
2454222.77671 & 0.00039 &  $-$277.0 & $ $0.00085 & \citet{Caceres+09aa} \\                                  
2454226.71808 & 0.00033 &  $-$276.0 & $ $0.00072 & This work (ISAAC data from \citealt{Caceres+09aa}) \\    
2454285.84036 & 0.00097 &  $-$261.0 & $ $0.00043 & C.\ Foote (AXA) \\                                       
2454506.56417 & 0.00010 &  $-$205.0 & $-$0.00003 & \citet{Burke+10apj} \\                                   
2454518.38906 & 0.00017 &  $-$202.0 & $ $0.00034 & \citet{Burke+10apj} \\                                   
2454553.86244 & 0.00100 &  $-$193.0 & $ $0.00018 & B.\ Gary (AXA) \\                                        
2454620.86554 & 0.00080 &  $-$176.0 & $-$0.00230 & B.\ Gary (AXA) \\                                        
2454620.86784 & 0.00080 &  $-$176.0 & $-$0.00000 & C.\ Foote (AXA) \\                                       
2454624.81004 & 0.00140 &  $-$175.0 & $ $0.00069 & C.\ Foote (AXA) \\                                       
2454624.81214 & 0.00130 &  $-$175.0 & $ $0.00279 & C.\ Foote (AXA) \\                                       
2454628.75154 & 0.00040 &  $-$174.0 & $ $0.00069 & Healy (AXA) \\                                           
2454888.89006 & 0.00070 &  $-$108.0 & $-$0.00012 & B.\ Gary (AXA) \\                                        
2454959.83746 & 0.00060 &   $-$90.0 & $ $0.00019 & B.\ Gary (AXA) \\                                        
2454959.83783 & 0.00150 &   $-$90.0 & $ $0.00056 & This work (data from \citealt{Sada+12pasp}) \\           
2454967.71916 & 0.00070 &   $-$88.0 & $-$0.00112 & B.\ Gary (AXA) \\                                        
2454983.48656 & 0.00080 &   $-$84.0 & $ $0.00026 & J.\ Gregorio (AXA) \\                                    
2454987.42836 & 0.00080 &   $-$83.0 & $ $0.00055 & Ayoinemas (AXA) \\                                       
2455058.37686 & 0.00100 &   $-$65.0 & $ $0.00196 & Srdoc (AXA) \\                                           
2455290.92347 & 0.00060 &    $-$6.0 & $-$0.00023 & B.\ Gary (AXA) \\                                        
2455298.80597 & 0.00060 &    $-$4.0 & $-$0.00074 & B.\ Gary (AXA) \\                                        
2455314.57290 & 0.00014 &       0.0 & $ $0.00017 & This work (INT/WFC light curve) \\                       
2455365.81217 & 0.00050 &      13.0 & $-$0.00013 & B.\ Gary (AXA) \\                                        
2455369.75357 & 0.00070 &      14.0 & $-$0.00023 & B.\ Gary (AXA) \\                                        
2455369.75517 & 0.00070 &      14.0 & $ $0.00137 & B.\ Gary (AXA) \\                                        
2455629.89263 & 0.00041 &      80.0 & $-$0.00052 & S.\ Shadic (TRESCA) \\                                   
2455653.54256 & 0.00056 &      86.0 & $ $0.00038 & R.\ Naves (TRESCA) \\                                    
2455700.84090 & 0.00053 &      98.0 & $ $0.00066 & S.\ Shadic \\                                            
2455712.66431 & 0.00077 &     101.0 & $-$0.00045 & S.\ Dvorak (TRESCA) \\                                   
2455834.85186 & 0.00017 &     132.0 & $ $0.00044 & \citet{Deming+13apj} \\                                  
2455984.62762 & 0.00061 &     170.0 & $-$0.00100 & J.\ Trnka (TRESCA) \\                                    
2456055.57528 & 0.00019 &     188.0 & $-$0.00044 & This work ($g$-band light curve from BUSCA) \\           
2456055.57529 & 0.00013 &     188.0 & $-$0.00043 & This work ($r$-band light curve from BUSCA) \\           
2456055.57614 & 0.00017 &     188.0 & $ $0.00042 & This work ($z$-band light curve from BUSCA) \\           
2456059.51669 & 0.00030 &     189.0 & $-$0.00053 & This work ($u$-band light curve from BUSCA) \\           
2456059.51659 & 0.00014 &     189.0 & $-$0.00063 & This work ($g$-band light curve from BUSCA) \\           
2456059.51756 & 0.00016 &     189.0 & $ $0.00034 & This work ($z$-band light curve from BUSCA) \\           
2456059.51881 & 0.00054 &     189.0 & $ $0.00159 & R.\ Naves (TRESCA) \\                                    
2456063.45989 & 0.00063 &     190.0 & $ $0.00116 & A.\ Carre\~no (TRESCA) \\                                
2456067.40080 & 0.00061 &     191.0 & $ $0.00057 & S.\ Poddan\'y (TRESCA) \\                                
2456106.81198 & 0.00105 &     201.0 & $-$0.00331 & S.\ Curry (TRESCA) \\                                    
2456106.81544 & 0.00127 &     201.0 & $ $0.00015 & D.\ Mitchell (TRESCA) \\                                 
\hline
\end{tabular} \end{center} \end{table*}
\begin{table*}\begin{center}\contcaption{}
\begin{tabular}{l l r r l} \hline
Time of minimum & Uncertainty & Cycle & Residual & Reference \\
(BJD/TDB)) & (d) & number & (d) &       \\
\hline
2456130.46286 & 0.00185 &     207.0 & $-$0.00146 & F.\ Emering (TRESCA) \\                                  
2456725.63020 & 0.00067 &     358.0 & $-$0.00145 & M.\ Zibar (TRESCA) \\                                    
2456729.57470 & 0.00039 &     359.0 & $ $0.00155 & M.\ Zibar (TRESCA) \\                                    
2456737.45381 & 0.00086 &     361.0 & $-$0.00235 & J.\ Trnka (TRESCA) \\                                    
2456800.52272 & 0.00073 &     377.0 & $ $0.00246 & CAAT (TRESCA) \\                                         
2456804.46170 & 0.00015 &     378.0 & $-$0.00006 & This work (CAHA 1.23m light curve) \\                    
2457210.43605 & 0.00097 &     481.0 & $-$0.00079 & J.\ Trnka (TRESCA) \\                                    
2457210.43733 & 0.00032 &     481.0 & $ $0.00049 & M.\ Bretton (TRESCA) \\                                  
2457257.73677 & 0.00130 &     493.0 & $ $0.00186 & O.\ Mazurenko (TRESCA) \\                                
2457454.80941 & 0.00125 &     543.0 & $-$0.00079 & K.\ Menzies (TRESCA) \\                                  
2457478.45983 & 0.00081 &     549.0 & $ $0.00060 & A.\ Marchini (TRESCA) \\                                 
2457545.46405 & 0.00101 &     566.0 & $-$0.00078 & F.\ Lomoz (TRESCA) \\                                    
2457545.46612 & 0.00083 &     566.0 & $ $0.00129 & F.\ Lomoz (TRESCA) \\                                    
\hline \end{tabular} \end{center} \end{table*}

\begin{figure*} \includegraphics[width=\textwidth,angle=0]{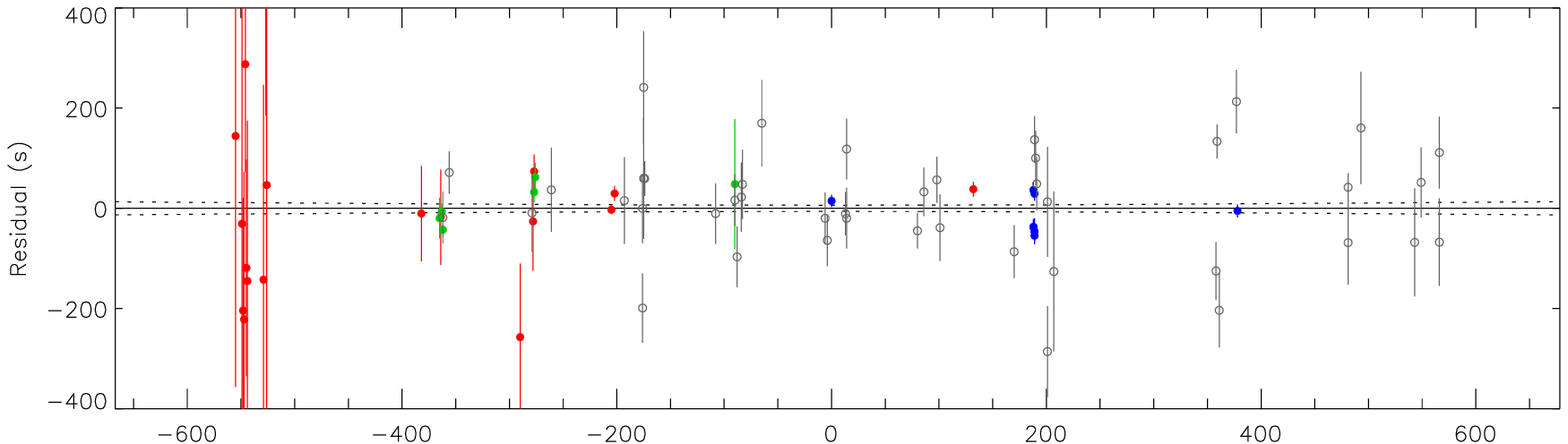}
\caption{\label{fig:tmin} Plot of the residuals of the timings of mid-transit
versus a linear ephemeris. The results from the new data in this work are shown
in blue, from published data reanalysed in this work in green, from published
papers in red, and from amateur observers in grey. The dotted lines show the
1$\sigma$ uncertainty in the ephemeris as a function of cycle number.} \end{figure*}

A crucial part of obtaining observations of XO-1 with JWST is the availability of a high-precision orbital ephemeris for the scheduling of observations. The most recent detailed study of the ephemeris of XO-1 is as long ago as that of \citet{Burke+10apj}. We have therefore redetermined times of minimum \reff{from} all available transit light curves in order to \reff{obtain} an ephemeris with the highest possible precision.

We first measured the times of mid-transit for each of our own light curves by fitting the data from each passband and each night with $T_0$, $r_{\rm A}+r_{\rm b}$, $k$, $i$, the linear LD coefficient of the quadratic law, and the relevant coefficients of the baseline polynomials as fitted parameters. All times of mid-transit are collected in Table\,\ref{tab:tmin}. The uncertainty in each measured $T_0$ was calculated using 1000 Monte Carlo simulations and residual-permutation simulations and the larger of the two errorbars kept.

We performed the same steps for the published light curves which we included in our analysis above. The photometry from some of these sources \citep{Holman+06apj,Caceres+09aa} is given on the ``HJD'' timescale, which we assumed to mean HJD/UTC and therefore converted into BJD/TDB for consistency with modern analysis methods. The data from \citet{Sada+12pasp} are already expressed as a function of BJD/TDB; however we found a large offset between our and their results which is probably due to the differing treatments of the out-of-transit baseline. Our measured $T_0$ has a significantly larger errorbar and also a better agreement with the final linear ephemeris.

\citet{Mccullough+06apj} quoted one time of mid-transit based on their follow-up photometry. \citet{Wilson+06pasp} presented two timings from the original XO survey data \citep{Mccullough+06apj} as well as nine times of mid-transit from SuperWASP data \citep{Pollacco+06pasp}. We ignored one timing with a quoted uncertainty of 31\,min. One more timing was obtained from \citet{Raetz+09an}. The timings discussed in this paragraph so far were quoted as being on the ``HJD'' system: we have assumed this to represent HJD/UTC and converted them all to BJD/TDB for consistency. Finally, we obtained two timings from \citet{Burke+10apj} and one from \citet{Deming+13apj}, all three being on the BJD/TDB timescale.

XO-1 was one of the earliest-discovered transiting planetary systems and has a deep transit well suited for observation with small telescopes. It therefore has a rich history of timings obtained by amateur observers. These have been systematically accumulated and fitted by contributors to the Exoplanet Transit Database (ETD\footnote{{\tt http://var2.astro.cz/ETD/credit.php}}; \citealt{Poddany++10newa}). We have included all timings based on observations of a complete transit with a scatter sufficiently low to clearly identify the transit shape by eye (sometimes by recourse to the AXA\footnote{{\tt http://brucegary.net/AXA/x.htm}} website), resulting in 43 $T_0$ values. These were all assumed to be on the HJD/UTC system and converted to BJD/TDB.

We fitted all times of mid-transit with a straight line to give the linear ephemeris:
$$ T_0 = {\rm BJD(TDB)} \,\, 2\,455\,314.572766 (49) \, + \, 3.94150514 (20) \times E $$
where the bracketed numbers show the uncertainty in the final digit of the preceding number and $E$ gives the cycle count versus the reference epoch. We chose the transit observed with the INT as the reference transit because it is close to the weighted mean of the $T_0$ values so the two terms in the ephemeris have a negligible correlation. The \chir\ of the fit is 1.66, a typical value for this kind of analysis \citep[e.g.][]{Me+16mn}. We interpret this as an indication that the errorbars of the individual measurements are modestly underestimated, and not as evidence of transit timing variations. We have multiplied the errorbars for the ephemeris by $\sqrt{1.66}$ to account for this -- the orbital period of the XO-1 system is now known to a precision of 0.017\,s. The residuals versus the linear ephemeris are shown in Fig.\,\ref{fig:tmin}.

\subsection{Constraints on orbital decay}

Tidal effects dominate the orbital evolution of short-period giant planets \citep[e.g.][]{Ogilvie14araa}. Tidally-induced orbital decay is expected to shorten the orbital period of XO-1 and shift its transits earlier in time \reff{in the usual case that the stellar rotation period exceeds the planet orbital period} \citep{Levrard++09apj,Jackson++09apj}. Tidal evolution timescales depend on the stellar tidal quality factor, $Q_\star$, which has a canonical value of $10^6$ but is uncertain by several orders of magnitude \citep{OgilvieLin07apj,Jackson++08apj2,PenevSasselov11apj,Penev+12apj}.

The relatively long observational history of XO-1 means that it is reasonable to check if transit times are useful in constraining the strength of $Q_\star$. Orbital decay would give rise to a progressive advance of the time of transit, imprinting a quadratic term in its orbital ephemeris. We fitted a quadratic ephemeris to the transit times collected in Table\,\ref{tab:tmin}, finding that the quadratic term was consistent with zero ($6.2 \times 10^{-10} \pm 9.0 \times 10^{-10}$\,d\,d$^{-1}$, or $9.7 \pm 14.2$\,ms\,yr$^{-1}$). The Bayesian Information Criterion \citep{Schwarz78} is higher for this ephemeris (219.6) than for the linear ephemeris (216.1). So is the Akaike Information Criterion \citep{Akaike81} with 212.7 versus 211.5, respectively. We conclude that there is no observational support for a quadratic ephemeris, and thus no detection of orbital decay in this planetary system.

To derive an upper limit on orbital decay, and thus a lower limit on $Q_\star$, we followed the procedure outlined by \citet{Birkby+14mn} and rediscussed by \citet{Wilkins+17apj}. In this method, the quadratic term in the orbital ephemeris, $q$, constrains the modified tidal quality factor
$$ Q_\star^{\,\prime} = \frac{3}{2} \, \frac{Q_\star}{k_2} $$
where $k_2$ is the Love number \citep{Love11book}. The relevant equation is\footnote{Note that the term $(\Porb/2\pi)$ is inverted in equations 3 and 5 of \citet{Wilkins+17apj}.}
$$ Q_\star^{\,\prime} = \frac{-27}{~~8} \left(\frac{M_{\rm b}}{M_{\rm A}}\right) \left(\frac{R_{\rm A}}{a}\right)^5 \left(\frac{\Porb}{2\pi}\right) \frac{1}{q} $$
The quantity $(R_{\rm A}/a)$ is of course the fractional radius of the star, $r_{\rm A}$, measured directly from the transit light curves in Section\,\ref{sec:lc}.

As the quadratic term is formally greater than zero -- which equates to an increasing orbital period -- we set the 3$\sigma$ limit on orbital decay to be $(q - 3\sigma_q) = -2.1 \times 10^{-9}$\,d\,d$^{-1}$ (i.e.\ $-33$\,ms\,yr$^{-1}$). Using the quantities in Table\,\ref{tab:absdim} and this constraint on $q$, we find a lower limit on the tidal quality factor to be $Q_\star^{\,\prime} > (4.0 \pm 0.3) \times 10^5$. The uncertainty was calculated by propagating the errors on $M_{\rm A}$, $M_{\rm b}$ and $r_{\rm A}$ with a Monte Carlo approach. For ease of comparison, this limit can also be expressed as $Q_\star^{\,\prime} > 10^{5.60\pm0.03}$.

\subsection{Constraints on periodic transit timing variations}

\begin{figure*} \includegraphics[width=\textwidth,angle=0]{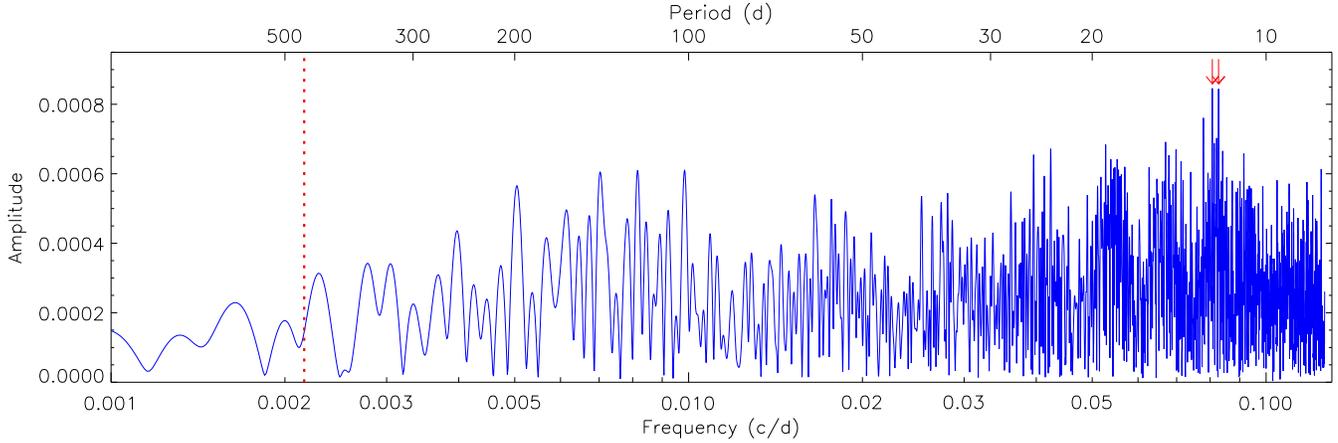}
\caption{\label{fig:pgram} Periodogram of the residuals of the timings of mid-transit
versus a linear ephemeris (blue solid line). The period of the tentative sinusoidal
variation found by \citet{Burke+10apj} is shown in a red dotted line. The two highest
peaks in the frequency spectrum are indicated using red arrows.} \end{figure*}

\citet{Burke+10apj} investigated a possible sinusoidal variation in the transit timing values with a period 118.3 orbital cycles, following a suggestion by B.\ Gary. They found that this more complex ephemeris provided a better fit to the data but by an amount which fell far short of statistical significance. To check this out we calculated a periodogram of the residuals of the best-fitting linear ephemeris with the {\sc period04} code \citep{LenzBreger04iaus} covering the frequency range from 0.0 to the Nyquist frequency of 0.13 cycles per day (i.e.\ equivalent to twice the orbital period).

Fig.\,\ref{fig:pgram} shows the resulting frequency spectrum. The red dotted line indicates the possible period at 118.3 orbital cycles (466.3\,d) mentioned by \citet{Burke+10apj}: the periodogram shows no significant power at this period. The two strongest peaks are at much higher frequencies of 0.0807 and 0.0802 cycles per day, and both have a signal to noise ratio of 3.04. This is well below the value of 4.0 typically considered to be the level at which a frequency is significant \citep[e.g.][]{Breger+93aa}. We therefore conclude that there is no evidence for a periodic variation in the orbital ephemeris of XO-1.



\section{The optical-infrared transmission spectrum of XO-1\,b}                                                                                                                         \label{sec:transspec}

\refff{We now} study how the transit depth varies as a function of wavelength. This effect is caused by changes in the apparent radius of the planet, which in turn arise from variations in opacity and scattering processes in its extended atmosphere. Its transmission spectrum therefore potentially holds information about the abundances of atoms and molecules, and the temperature structure of the atmosphere.

Following the approach of \citet{Me+12mn2}, we modelled all available transit light curves of XO-1 in order to measure the planet radius (in the form of $r_{\rm b}$) as a function of wavelength. It is important to fix the geometric parameters to representative values in order to maximise the consistency between different light curve fits and to avoid sources of uncertainty which are common to all light curves. The choice of these parameters is not simple because of conflicting results from published transmission spectroscopic studies of XO-1\,b.

\subsection{Consideration of published results}

\begin{figure*}
\includegraphics[width=\textwidth,angle=0]{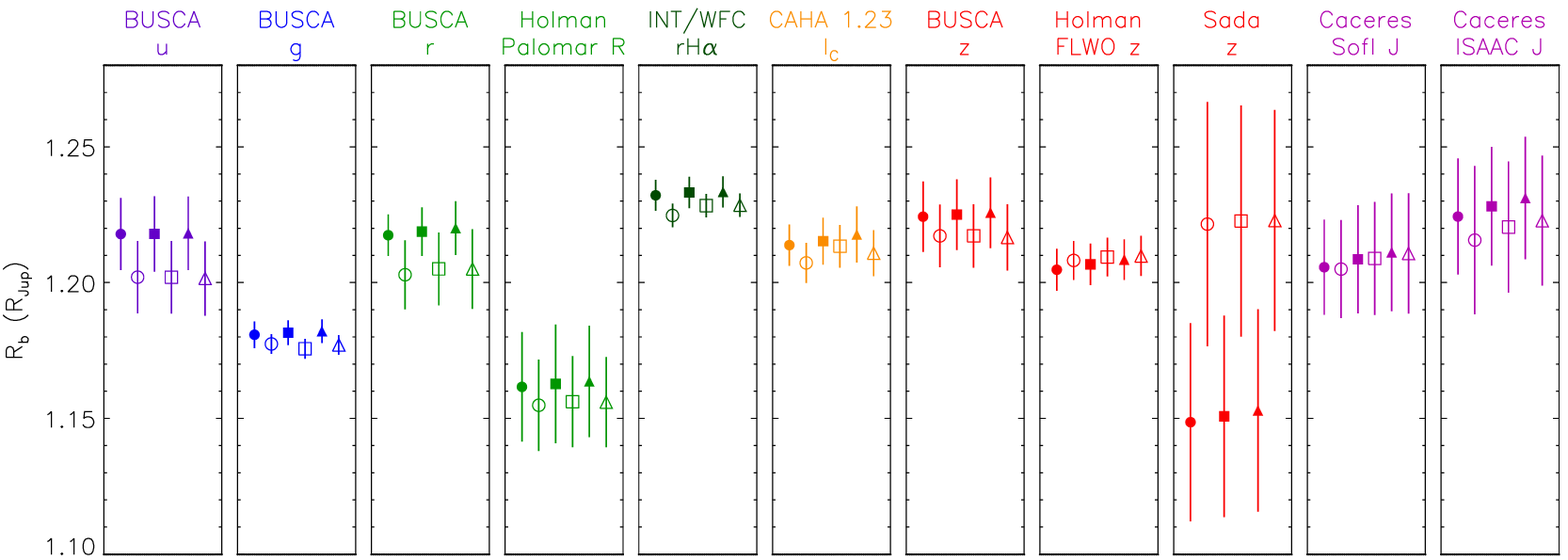}
\includegraphics[width=\textwidth,angle=0]{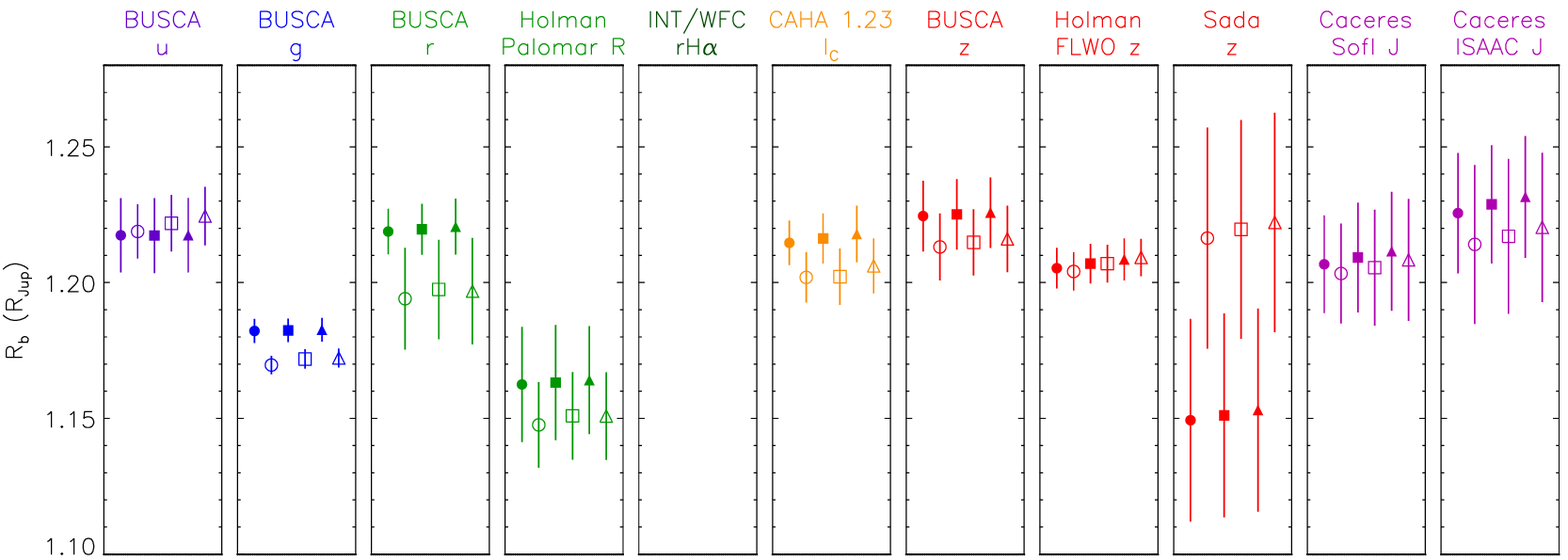}
\caption{\label{fig:ts:r2ld} Plot of the planet radius measured from each light curve for multiple alternative treatments of LD.
Filled symbols refer to measurements with one LD coefficient fitted and one fixed, and open symbols to value obtained when both
LD coefficients were fixed. Circles indicate the quadratic law, squares the logarithmic law, and triangles the square-root law.
The LD coefficients were from {\sc atlas9} (upper panels) or {\sc phoenix} (lower panels) model atmospheres. The source data and
passband are specified at the top of each panel and colour coding is as Figs.\ \ref{fig:lcall} and \ref{fig:lcfit}.} \end{figure*}

\citet{Tinetti+10apj} presented HST/NICMOS observations of a transit of XO-1 which yielded a transmission spectrum covering 1.2--1.8\,$\mu$m. They claimed the detection of H$_2$O, CH$_4$ and CO$_2$ molecules in the planetary atmosphere. \citet{Burke+10apj} extended this analysis to the geometric parameters of the system, and included a second (or should that be first?) NICMOS observation of XO-1 obtained 12\,days (three planetary orbits) prior to the observations utilised by \citet{Tinetti+10apj}.

\citet{Gibson++11mn} presented a reanalysis of the NICMOS data used by \citet{Tinetti+10apj}, with differences of approach concerning the use of decorrelation parameters to remove systematic errors in the data which arise from both HST and NICMOS. \citet{Gibson++11mn} obtained a more scattered and much more uncertain transmission spectrum, and concluded that the detection of molecules claimed by \citet{Tinetti+10apj} was not supported by the data. \citet{Gibson++11mn} concluded that NICMOS is not a suitable instrument for transmission spectroscopy as it displays unremovable systematics of similar size to the astrophysical signal being sought.

\citet{Crouzet+12apj} also presented a reanalysis of the NICMOS observations from \citet{Tinetti+10apj}, but also included the second transit of XO-1 observed 12\,days earlier. They performed a similar reduction of the data as \citet{Tinetti+10apj} and \citet{Gibson++11mn}, but with some different choices of instrumental parameters against which the light curves were decorrelated. They found results which were much closer to those of \citet{Tinetti+10apj} than \citet{Gibson++11mn}, but with important differences remaining at the level of the expected astrophysical signal in the transmission spectrum.

\citet{Deming+13apj} used the improved capabilities of HST/WFC3 to obtain a transmission spectrum of XO-1\,b over the 1.12--1.65\,$\mu$m wavelength interval. This was used to claim a detection of water absorption in the planetary atmosphere, as well as to rule out spectral features at the level claimed by \citet{Tinetti+10apj}. As the work by \citet{Deming+13apj} is based on a more modern analysis of data obtained using a better instrument than previous transmission spectroscopy, we have chosen to anchor our new results on the geometric parameters used in this work. They are, in turn, those found by \citet{Burke+10apj}: $r_{\rm A} = 0.0890 \pm 0.0007$ (the inverse of the quoted quantity $\frac{a}{R_\star} = 11.24 \pm 0.09$) and $i = 88.8 \pm 0.2^\circ$.

\subsection{Analysis method}

For each light curve we calculated the best-fitting model with {\sc jktebop}. We fixed $r_{\rm A}$ at 0.0890, $i$ at 88.8$^\circ$ and the orbital period at a representative value. We fitted for $r_{\rm b}$, the time of mid-transit (to guard against possible orbital period variations) and the coefficients of the baseline polynomial \reff{(see Table\,\ref{tab:obslog})}. Uncertainties in $r_{\rm b}$ were calculated using both Monte Carlo and residual-permutation simulations, and the larger errorbar for $r_{\rm b}$ was retained in each case. We found that the uncertainties for the BUSCA $z$-band light curve were relatively large, especially for the residual-permutation simulations: this is a result of the moderate differences between the two light curves and therefore is expected.

The phenomenon of LD deserves special consideration. In a recent work on GJ\,1132 \citep{Me+17aj}, and in provisional analyses for the current work, we found that the transmission spectrum was significantly affected by way in which LD was treated. We therefore modelled the light curves with a range of ways of dealing with LD. The quadratic LD law is the most widely used in the literature, but recent theoretical studies \citep{EspinozaJordan16mn,Morello+17aj} have found that other laws, such as logarithmic and square-root \citep[see][for the equations]{Me08mn}, are capable of matching theoretical LD predictions more precisely. Logarithmic should be better than square-root in the current case, particularly for the redder optical passbands under consideration \citep{Vanhamme93aj}.

We therefore obtained solutions to the light curves using the quadratic, logarithmic and square-root LD laws, in each case with both coefficients fixed and with the linear coefficient fitted but the nonlinear coefficient fixed. For consistency we adopted theoretical LD coefficients obtained by \citet{Claret00aa,Claret04aa2} using the {\sc atlas9} atmosphere models \citep{Kurucz93}, for all light curves, with the exception of the redshifted H$\alpha$ filter for which we used LD coefficients from the Johnson $R$ filter tabulated by \citet{Vanhamme93aj}.

For a comparison with the results above, and in order to capture the effect of differences in the LD coefficients used, we also fit each light curve using LD coefficients from \citet{Claret00aa,Claret04aa2} calculated using the {\sc phoenix} model atmospheres. Fig.\,\ref{fig:ts:r2ld} shows the results for all alternatives investigated. It can be seen that
the measured value of $r_{\rm b}$ is {\em not} significantly affected by either the choice of LD law, whether or not one of the LD coefficients is fitted, or whether the LD coefficients come from the {\sc atlas9} or {\sc phoenix} model atmospheres. We also notice -- perhaps counterintuitively -- that fixing both LD coefficients can yield larger errorbars despite the loss of one dimension from the area of parameter space in which the solution can be located. This occurs because fixing the LD coefficients can cause a poorer fit to the data, leading to larger errorbars from the residual-permutation algorithm.

From Fig.\,\ref{fig:ts:r2ld} we conclude that the treatment of LD does not have a significant effect on the results for individual light curves, and that it is safe to proceed with a representative set of $r_{\rm b}$ measurements. One possible exception to this rule is the $g$-band, for which the effect of LD treatment on the measured planet radius is significantly above the (very small) errorbars. Notwithstanding this, we \reff{chose} as the representative set of $r_{\rm b}$ values those measured using the quadratic LD law with the linear coefficient fitted at values from the {\sc atlas9} model atmospheres. Table\,\ref{tab:rb} contains these values, and also for reference contains those from the quadratic LD law with both LD coefficients fixed. Table\,\ref{tab:rb} also includes values for the central wavelength and full width at half maximum of the filters used to obtain our observations with BUSCA\footnote{{\tt https://www.caha.es/CAHA/Instruments/filterlist.html}} and the INT/WFC\footnote{{\tt http://catserver.ing.iac.es/filter/list.php?\\instrument=WFC}}, and for published data obtained using the Palomar 50\,in\footnote{{\tt http://www.astro.caltech.edu/palomar/observer/\\60inchResources/p60filters.html}}, SofI\footnote{{\tt http://www.eso.org/sci/facilities/lasilla/\\instruments/sofi/inst/Imaging.html}} and ISAAC\footnote{{\tt http://www.eso.org/sci/facilities/paranal/\\decommissioned/isaac/doc/VLT-MAN-ESO-14100-0841\_v90.pdf}} instruments.

\subsection{Results}

\begin{table*} \centering
\caption{\label{tab:rb} Values of $r_{\rm b}$ for each light curve. The errorbars in this table
exclude all common sources of uncertainty so should only be used to interpret relative differences
in $r_{\rm b}$. The central wavelengths and full widths at half maximum transmission are given for
the filters used to obtain our own data. Values of $r_{\rm b}$ are given for two cases: both LD
coefficients fixed, and the linear LD coefficient fitted but the quadratic LD coefficient fixed.
In both cases the quadratic LD law was used and LD coefficients came from the {\sc atlas9} model
atmospheres.}
\begin{tabular}{l l r r r@{\,$\pm$\,}l r@{\,$\pm$\,}l} \hline
Data    & Filter & Central         & Band full  & \mc{$r_{\rm b}$} & \mc{$r_{\rm b}$} \\
source &         & wavelength (nm) & width (nm) & \mc{(LD fixed} & \mc{(LD fitted)}   \\
\hline
INT/WFC         & redshifted H$\alpha$ &  689 &  10 & 0.011983 & \refff{0.000210} & 0.011911 & \refff{0.000150} \\
BUSCA           & SDSS $u$             &  366 &  38 & 0.011844 & 0.000129 & 0.011690 & 0.000129 \\
BUSCA           & SDSS $g$             &  478 & 150 & 0.011483 & 0.000048 & 0.011450 & 0.000035 \\
BUSCA           & SDSS $r$             &  663 & 105 & 0.011840 & 0.000074 & 0.011698 & 0.000123 \\
BUSCA           & SDSS $z$             &  910 &  90 & 0.011906 & 0.000126 & 0.011837 & 0.000112 \\
CAHA 1.23\,m    & Cousins $I$          &  810 & 110 & 0.011804 & 0.000074 & 0.011741 & 0.000071 \\
Holman FLWO     & SDSS $z$             &      &     & 0.011716 & 0.000075 & 0.011750 & 0.000069 \\
Holman Palomar  & Cousins $R$          &  647 & 152 & 0.011297 & 0.000196 & 0.011231 & 0.000164 \\
C\'aceres SofI  & $J$                  & 1247 & 290 & 0.011725 & 0.000170 & 0.011719 & 0.000175 \\
C\'aceres ISAAC & $J$ + block          & 1250 & 290 & 0.011907 & 0.000208 & 0.011822 & 0.000265 \\
Sada            & $z^\prime$           &      &     & 0.011170 & 0.000355 & 0.011880 & 0.000437 \\
\hline \end{tabular} \end{table*}

\begin{figure} \includegraphics[width=\columnwidth,angle=0]{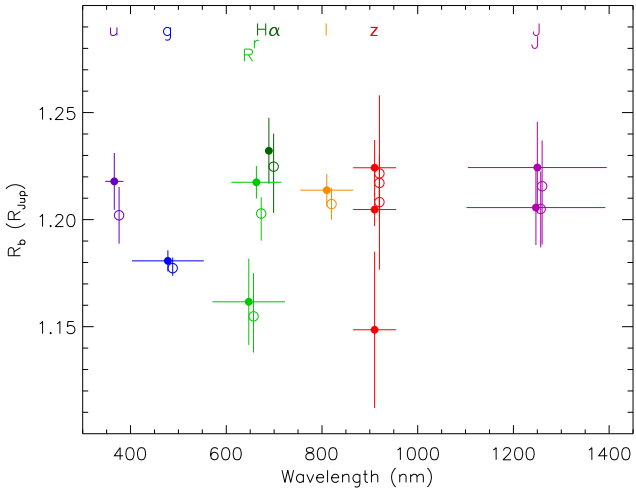}
\caption{\label{fig:rvary} \refff{Measured planetary radius ($R_{\rm b}$) as a function
of the central wavelength of the passbands used. The passband names are give at
the top of the plot. The horizontal lines indicate the FWHM of the passband used
and the vertical lines show the errorbars in the $R_{\rm b}$ measurements. {\em The
errorbars exclude all common sources of uncertainty}. Results obtained when fitting
the linear LD coefficient are shown as filled circles with errorbars. Results from
fixing both LD coefficients are shown as open circles without a horizontal line
indicating the passband. The colour coding is consistent with Figs.\ \ref{fig:lcall}
and \ref{fig:lcfit}.}} \end{figure}

In Fig.\,\ref{fig:rvary} we show the transmission spectrum of XO-1\,b determined from the light curves studied in this work, both new and previously published. Our preferred approach is to fit for the linear LD coefficient, and these results are shown as filled circles. The alternative approach of fixing both LD coefficients yields the results shown using open circles. Fig.\,\ref{fig:rvary} shows the values of $R_{\rm b}$ obtained by multiplying the $r_{\rm b}$ values in Table\,\ref{tab:rb} by the semimajor axis (0.04914\,AU) and a conversion factor (1\,AU $=$ 2092.5\Rjup).

It is immediately apparent from Fig.\,\ref{fig:rvary} that different light curves in the same or similar passbands show significant \reff{variations in $r_{\rm b}$}. On closer inspection the two worst offenders are the Palomar $R$-band data from \citet{Holman+06apj} and the $z$-band light curve from \citet{Sada+12pasp}. Both have a high scatter and include no observations on one side of the transit, so it is not surprising that they give $r_{\rm b}$ values which are very uncertain. This issue can be dealt with either by combining results from multiple light curves in the same or similar passbands or by ignoring the problematic results. In the current case, both options give a similar outcome.


\begin{figure} \includegraphics[width=\columnwidth,angle=0]{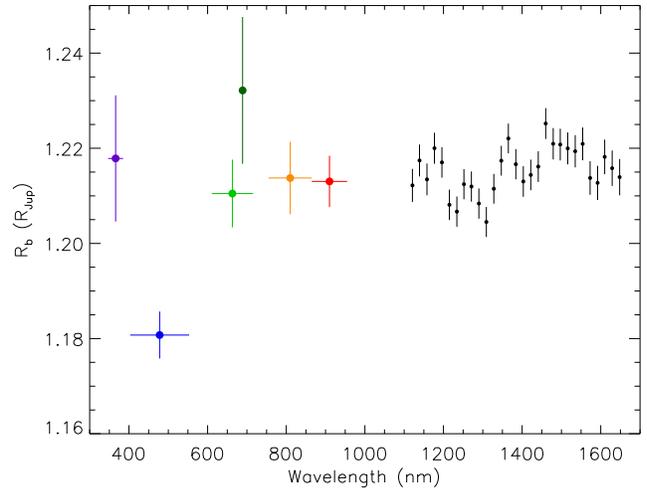}
\caption{\label{fig:rvary2} \refff{As Fig.\,\ref{fig:rvary} but similar passbands have
been combined into weighted mean values (the $z$-bands and the $R$ band and $r$ band),
and the $J$-band results are ignored in favour of the HST/WFC3 transmission spectrum
of XO-1\,b obtained by \citet{Deming+13apj} and shown using black filled circles.}} \end{figure}

In Fig.\,\ref{fig:rvary2} we show the transmission spectrum of XO-1\,b after some consolidation of the results. The three $z$-band $r_{\rm b}$ values have been reduced into their weighted mean, as have the Palomar $R$ and BUSCA $r$ bands, in order to stop their large errorbars obfuscating such plots. We have not combined the redshifted H$\alpha$ result with any other as the value of $r_{\rm b}$ from this light curve has \reff{much greater wavelength resolution (resolving power $R \approx 70$) than the $R$ and $r$ bands}. We have furthermore ignored the $J$-band results from now on because they add nothing to our analysis: they are consistent with and are completely overlapped by published transmission spectra, but are of \reff{lower precision and much lower wavelength resolution}.

In Fig.\,\ref{fig:rvary2} we have also plotted the HST/WFC3 transmission spectrum of XO-1\,b obtained by \citet{Deming+13apj}, after converting it from the values of $k^2$ \citep[][their table\,3]{Deming+13apj} to $R_{\rm b}$ consistently with our values of $r_{\rm b}$. The treatment of LD by \citet{Deming+13apj} is relevant: they used the linear LD law with coefficients fixed to values interpolated from the $J$- and $H$-band coefficients tabulated by \citet{ClaretBloemen11aa}. They account for minor variations between different sources of theoretical LD coefficients, but do not allow for any imperfections in the description of real stars by current theoretical model atmospheres. They also neglect the spectral variation of LD coefficients over wavelength intervals smaller than those of the broad-band $J$ and $H$ filters. This approach is quite simplistic, but has less impact in the infrared than at visual wavelengths, because stellar LD is weaker in the infrared.

\subsection{Interpretation}

\begin{figure} \includegraphics[width=\columnwidth,angle=0]{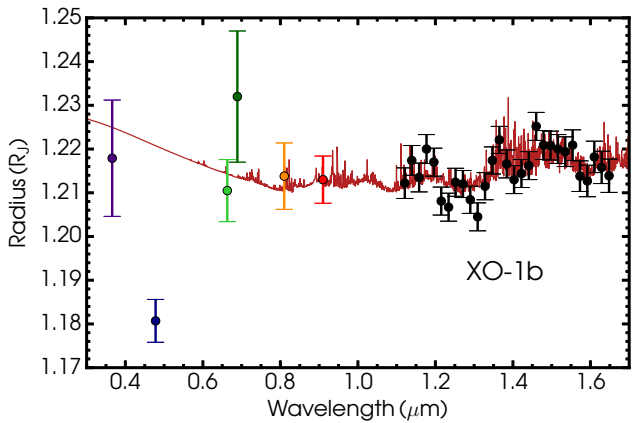}
\caption{\label{fig:madhu} Best-fitting model transmission spectrum of XO-1\,b
(dark red line). The observed transmission spectrum is shown using coloured points
for the optical data and black points for the HST/WFC near-infrared data.} \end{figure}

We used a forward transmission spectrum model to fit the optical and near infrared data of XO-1\,b. For the pressure-temperature profile, we use the parameterisation of \citet{MadhusudhanSeager09apj} which consists of six free parameters. We partitioned our model atmosphere into 100 layers spaced equally in log-pressure between $10^{-6}$ bar and $10^{2}$ bar. For the atmospheric composition, we considered several chemical species with prevalent signatures in the the spectral range of the optical and near-infrared observations \citep{Madhusudhan12apj,Moses+13apj,Venot+13aa}. These include Na, K, H$_2$O, NH$_3$, HCN, and CH$_4$. The mixing ratio of each species was assumed to be uniform in the observable atmosphere and we assumed an atmosphere rich in H$_2$ and He with a He/H$_2$ ratio of 0.17. We considered line absorption from each molecular species and collision-induced opacity from H$_2$-H$_2$ and H$_2$-He. The sources of opacity for the chemical species are described in \citet{GandhiMadhusudhan17mn,GandhiMadhusudhan18mn}. In addition, we accounted for cloud effects due to small and large modal particle sizes. Large cloud particles were represented by a grey opacity throughout the whole spectrum and small cloud particles and/or hazes modified the H$_2$ scattering Rayleigh slope in the optical.

The full set of observations were best fitted (Fig.\,\ref{fig:madhu}) with a patchy cloud model having a terminator cloud and haze fraction of 0.54. The patchy cloud model is generally preferred to a clear-atmosphere model at the 1.3$\sigma$ confidence level. H$_2$O is present at 3.05$\sigma$ confidence to fit the HST/WFC3 data, signifying water vapour is present with a certainty of 99.87\%. Nitrogen chemistry (NH$_3$ and HCN) is hinted at 1.5$\sigma$. The data do not provide evidence for the presence of either Na or K in the planetary atmosphere. Our model fits the optical transmission spectrum in the $u$, $r$/$R$, $i$ and $z$ bands to within 0.5$\sigma$.



The best-fit model is unable to explain the measured planet radius in the $g$ band, which lies 8$\sigma$ below the model transmission spectrum and well below all other planet radius measurements. The reason for this discrepancy is not clear but is very difficult to explain theoretically, as none of our model transmission spectra exhibit a planet radius at any point in the optical which is below the radii in the infrared. It is also hard to understand observationally, as the two light curves in this passband are of high precision and very good mutual agreement, and such an effect has not been seen in this band in previous observations by our team\footnote{For example WASP-57 \citep{Me+15mn2}, HAT-P-23 and WASP-48 \citep{Ciceri+15aa2}, Qatar-2 \citep{Mancini+14mn} and HAT-P-32 \citep{Tregloan+18mn}.}. Temporal variability of the planet \reff{or stellar  \citep[e.g.][]{Oshagh+14aa,Rackham+17apj} atmosphere} cannot be culpable because both $g$-band light curves were obtained simultaneously with $z$-band and either $u$-band or $r$-band observations.


We conclude that the transmission spectrum is best reproduced by a H$_2$/He-rich planetary atmosphere containing H$_2$O with low confidence levels of patchy clouds and nitrogen-bearing molecules (NH$_3$ and HCN). An anomalously small planet radius in the $g$-band is difficult to explain either observationally or theoretically and should be investigated by obtaining new observations in this wavelength region, preferably with a significantly higher resolution. 

\subsection{Discrepant transit depths}

\reffff{The referee expressed concern over the discrepant transit depth obtained from the $g$-band light curves. It is clear that there is something affecting the $g$-band data which is not accounted for in our data reduction and analysis procedures. These datasets were processed through the same data reduction and analysis programs as used by our group in many previous studies, which implies that the problem lies with the data themselves rather than with the reduction and analysis. Based on this, we rejected the $g$-band data from the analysis of the transmission spectrum. This implicitly assumes that the problem is isolated to the $g$-band alone; our results could be affected if the problem exists in other light curves or is an artefact of our data reduction pipeline.}

\reffff{We chose not to reject the $g$-band data when determining the physical properties of the system, and have assessed the impact of this choice by rerunning the analysis without the $g$-band data. We find that the final photometric results (Table\,\ref{tab:lcfit}) differ by 0.3$\sigma$ for $r_{\rm b}$ and less than 0.1$\sigma$ for $i$ and $r_{\rm A}$. The physical properties of the system in Table\,\ref{tab:absdim} are unchanged except that $R_{\rm b}$ increases by 0.2$\sigma$ and $\rho_{\rm b}$ decreases by 0.3$\sigma$. The inclusion of the $g$-band data therefore does not have a significant effect on the measured physical properties of the XO-1 system.}

\subsection{Impact of the optical data}

\begin{figure*} \includegraphics[width=\textwidth,angle=0]{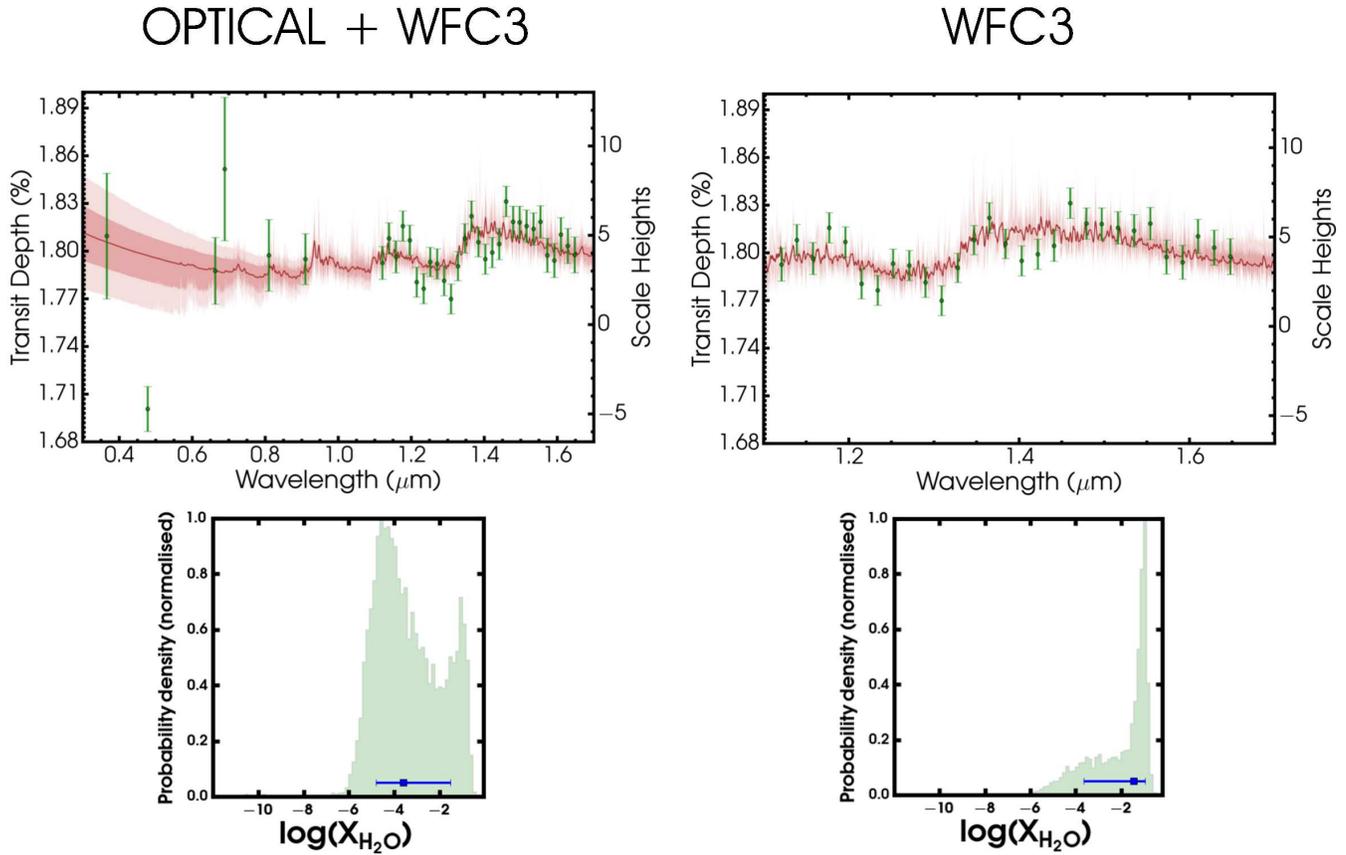}
\caption{\label{fig:OPTICAL+WFC3} \refff{Retrieved model transmission spectra of XO-1\,b
observations for the optical and near-infrared (left) and near-infrared only (right).
The observations are shown in green and the retrieved median model is in dark red
with associated 1$\sigma$ and 2$\sigma$ confidence contours. The median model in dark
red has been smoothed for clarity. The probability density function of the water
abundance is shown in the lower panels for both cases, where the points and errors
represent the median abundance and 1$\sigma$ intervals, respectively.}} \end{figure*}

\reff{One purpose of the current work was to see what improvement in our understanding of the properties of the atmosphere of XO-1\,b could be obtained by adding optical transit data to the HST near-infrared transmission spectrum. We investigated this by modelling both the full transmission spectrum and the HST results only.}

\reff{We find that the addition of the optical data to the near-infrared observations introduces an alternative water abundance estimate. Fig.\,\ref{fig:OPTICAL+WFC3} shows the retrieved water abundances for the case of our optical observations plus the HST data, and for the HST data alone. In the latter case the modal H$_2$O abundance is approximately $-1$\,dex with a median and $1\sigma$ errorbars of $-1.45^{+0.50}_{-2.19}$. The slight tail of the posterior distribution arises from a weak degeneracy with HCN. The adjoined observations in the visible offer a complementary interpretation of XO-1\,b's atmosphere, adding a second mode to the H$_2$O mixing ratio at $-4$\,dex and thus altering the median abundance by approximately $-2$\,dex.}

\reff{The two interpretations of XO-1\,b's atmospheric H$_2$O concentration emerge from two possible cloud condensate configurations. The \refff{water abundance mode at
approximately $-4$\,dex that is introduced by the optical data suggests} an atmosphere with condensate clouds composed of particle sizes $\sim$1 $\mu$m whose cloud-top pressures are 0.01 to 0.1 mbar. The formation efficiency of condensate particles decreases with atmospheric height \citep{Parmentier++13aa}, and therefore clouds extending to low pressures of 0.01--0.1\,mbar require vertical mixing processes such as convection which could advect material upward. \refff{On the other hand, the second mode constituting a high water abundance of approximately $-1$\,dex proposes cloud-top pressures greater than 1\,mbar}. Ultimately, elucidating the atmosphere of XO-1\,b from these two distinct possibilities (low water abundance/high-extending clouds, and high water abundance/low-extending clouds) will have to await more precise observations in the optical.}


\section{Summary and discussion}
\label{sec:summary}

\reff{XO-1 has been identified as a good candidate for the JWST Early Release Science program \citep{Stevenson+16pasp}. A near-infrared transmission spectrum for XO-1\,b has previously been obtained using HST/WFC3, resulting in the detection of water in the planetary atmosphere.} We have obtained a total of ten high-precision transit light curves covering the full optical wavelength range (366\,nm to 910\,nm) in order to extend this transmission spectrum to optical wavelengths.

We use our data, alongside published transit light curves and spectroscopic quantities of the host star, to measure the physical properties of the system. Our results are in good agreement with, and more precise than, previous studies. We also assemble all available transit timing measurements and derive a high-precision orbital ephemeris useful for scheduling future observations. We find no evidence for periodic deviations from this ephemeris, contrary to previous suggestions. The non-detection of any quadratic deviation from the linear ephemeris allows us to constrain the tidal quality factor for the host star to be $Q_\star^{\,\prime} > 10^{5.60}$.

We fitted the transit light curves using the same system geometry as for the HST/WFC3 observations, in order to measure the radius of the planet as a function of wavelength. This optical-infrared transmission spectrum is well fitted by a model spectrum for a planet with a H$_2$/He-rich atmosphere and patchy cloud. H$_2$O is detected to 3.05$\sigma$ while suggestions of patchy clouds (1.3$\sigma$) and nitrogen chemistry (1.5$\sigma$) are weak given the present observations. \reff{We find that adding the optical to the near-infrared data leads to {\em less} precise constraints on the planetary atmosphere. This indicates that optical observations of a higher precision and spectral resolution would be needed to improve our understanding of the atmosphere of XO-1\,b, and also that there is some tension between the best-fitting atmospheric properties in the optical and in the near-infrared.} The planet radius we measure in the $g$-band is anomalously low, a finding difficult to explain either observationally or theoretically. We advocate further observations in this wavelength region, with a higher spectral resolution. 


Throughout this work we have paid careful attention to the treatment of LD when fitting transit light curves. When measuring the physical properties of the system we used four different LD laws and two different approaches to fitting the coefficients of these. We find that the range of solutions produced by these different fits is very small when fitting high-quality data, so the treatment of LD is thankfully not a significant hindrance to measuring the system properties. From a similarly detailed investigation concerning the transmission spectrum, we find that the choice of LD law, and whether or not to fit for one of the coefficients, is unimportant, giving rise to a scatter in the planet radius measurements which is small compared to the variation between light curves. The only exception to this rule is for the $g$-band, where the very small uncertainties in the planet radius do not fully cover the scatter between solutions with a different treatment of LD. Whilst the situation for XO-1 is encouraging, we urge that similar analysis should be performed as standard procedure when obtaining transmission spectra. This is particularly true for planets transiting low-mass stars, whose LD may not be well captured by parametric laws and for which LD coefficients are more difficult to derive theoretically.

We confirm that XO-1 is an excellent target for future observations with JWST. Its physical properties are well-understood, the planet's transmission spectrum has features comparatively easy to measure using existing instrumentation, its solar-type host star shows no sign of chromospheric activity, and our new orbital ephemeris is precise enough to predict transits to within $\pm$5\,s up to the year 2266.


\section*{Acknowledgements}

We thank the referee for comments which improved the paper and encouraged us to test the reliability of our results.
The reduced light curves presented in this work will be made available at the CDS ({\tt http://vizier.u-strasbg.fr/}) and at {\tt http://www.astro.keele.ac.uk/jkt/}.
JS acknowledges financial support from the Leverhulme Trust in the form of a Philip Leverhulme Prize. AP is grateful for research funding from the Gates Cambridge Trust. LM acknowledges support from the Italian Minister of Instruction, University and Research (MIUR) through FFABR 2017 fund, and from the Department of Physics of the University of Rome Tor Vergata, through Mission Sustainability 2016 funds. The following internet-based resources were used in research for this paper: the ESO Digitized Sky Survey; the NASA Astrophysics Data System; the SIMBAD database and VizieR catalogue access tool operated at CDS, Strasbourg, France; and the ar$\chi$iv scientific paper preprint service operated by Cornell University. Based on observations collected at the Centro Astron\'omico Hispano Alem\'an (CAHA) at Calar Alto, Spain, operated jointly by the Max-Planck Institut f\"ur Astronomie and the Instituto de Astrof\'{\i}sica de Andaluc\'{\i}a (CSIC), and on observations made with the Isaac Newton Telescope operated on the island of La Palma by the Isaac Newton Group in the Spanish Observatorio del Roque de los Muchachos of the Instituto de Astrof\'{\i}sica de Canarias


\bibliographystyle{mn_new}

\end{document}